\documentclass[12pt]{article}

\usepackage{graphicx}
\usepackage{fancybox}
\usepackage{amsmath}
\usepackage{amssymb}
\usepackage{latexsym}
\usepackage{epsfig}

\newcommand{\al}{\alpha}
\newcommand{\ep}{\epsilon}
\newcommand{\vep}{\varepsilon}
\newcommand{\bep}{\bm{\epsilon}}

\newcommand{\la}{\lambda}

\newcommand{\NS}{\mbox{NS}}
\newcommand{\tNS}{\widetilde{\mbox{NS}}}
\newcommand{\R}{\mbox{R}}
\newcommand{\tR}{\widetilde{\mbox{R}}}
\newcommand{\sNS}{\msc{NS}}

\newcommand{\del}{\partial}

\newcommand{\lb}{\lbrack}
\newcommand{\rb}{\rbrack}

\newcommand{\msc}[1]{\mbox{\scriptsize #1}}
\newcommand{\dsp}{\displaystyle}

\newcommand{\bc}{\Bbb C}
\newcommand{\br}{\Bbb R}
\newcommand{\bz}{\Bbb Z}

\newcommand{\bh}{{\Bbb H}}

\newcommand{\bm}[1]{\mbox{\boldmath ${#1}$}}

\newcommand{\cN}{{\cal N}}

\newcommand{\cF}{{\cal F}}

\newcommand{\cS}{{\cal S}}

\newcommand{\cQ}{{\cal Q}}

\newcommand{\cR}{{\cal R}}

\newcommand{\ta}{\tilde{a}}

\newcommand{\tnu}{\tilde{\nu}}

\newcommand{\hc}{\hat{c}}

\newcommand{\hZ}{\widehat{Z}}

\newcommand{\Th}[2]{\Theta_{#1,#2}}
\renewcommand{\th}{{\theta}}

\newcommand{\ch}[2]{\mbox{ch}^{#1}_{#2}}

\newcommand{\chd}{\mbox{ch}_{\msc{\bf dis}}}

\newcommand{\hchd}{\widehat{\mbox{ch}}_{\msc{\bf dis}}}
\newcommand{\hchds}[1]{\widehat{\mbox{ch}}_{\msc{\bf dis}}^{(#1)}}

\newcommand{\chics}[1]{{\chi^{(#1)}_{\msc{\bf con}}}}

\newcommand{\chids}[1]{{\chi^{(#1)}_{\msc{\bf dis}}}}

\newcommand{\hchids}[1]{\widehat{\chi}_{\msc{\bf dis}}^{(#1)}}

\newcommand{\erf}{\mbox{Erf}}
\newcommand{\erfc}{\mbox{Erfc}}
\newcommand{\sgn}{\mbox{sgn}}

\newcommand{\nn}{\nonumber\\}

\newcommand{\dis}{\bf dis}
\newcommand{\con}{\bf con}

\newcommand{\reg}{\bf reg}
\newcommand{\asp}{\bf asp}
\newcommand{\fin}{\bf fin}

\renewcommand{\Im}{{\rm Im}}

\newcommand{\any}{{}^{\forall}}

\renewcommand{\mod}{\, \mbox{mod} ~ }

\newcommand {\eqn}[1]{(\ref{#1})}

\makeatletter
\@addtoreset{equation}{section}
\def\theequation{\thesection.\arabic{equation}}
\makeatother

\begin{document}

%%% Title page %%%%%
\begin{titlepage}
 \
 \renewcommand{\thefootnote}{\fnsymbol{footnote}}
 \font\csc=cmcsc10 scaled\magstep1
 {\baselineskip=14pt
 \rightline{
 \vbox{\hbox{August, 2012}
       }}}

 \baselineskip=20pt
%\vskip 1cm
 
\begin{center}

{\bf \Large  Thermodynamics of Superstring on Near-extremal NS5 \\
and \\
Effective Hagedorn Behavior
} 

 \vskip 1.2cm
 
\noindent{ \large Yuji Sugawara}\footnote{\sf ysugawa@se.ritsumei.ac.jp}
\\
%{\sf ysugawa@se.ritsumei.ac.jp}

\medskip

 {\it Department of Physical Science, 
 College of Science and Engineering, \\ 
Ritsumeikan University,  
%\\
%1-1-1 Noji-Higashi, 
%Kusatsu
Shiga 525-8577, Japan}
 
%\vskip 2mm

\end{center}

\bigskip

\begin{abstract}

We study the thermodynamical torus partition function of superstring 
on the near-extremal black NS5-brane background. 
The exact partition function has been computed with the helps of our previous works:
[arXiv:1012.5721 [hep-th]], [arXiv:1109.3365 [hep-th]], 
and  naturally decomposed into two parts. 
The first part is contributed from strings freely propagating 
in the asymptotic region, which are identified as the 
superstring gas at the Hawking temperature on the linear-dilaton background.
The second part includes the contribution localized around the `tip of cigar', 
which characterizes the non-extremality. Remarkably, the latter part includes 
massless excitations with non-vanishing thermal winding, which signifies 
that the Hagedorn-like behavior effectively appears, even though 
the Hawking temperature is much lower than the Hagedorn temperature. 
We also explore the high-temperature backgrounds defined by 
the orbifolding along the Euclidean time direction. 
In those cases, the thermal winding modes localized around the tip are found to be 
tachyonic, reflecting the singularities  of Euclidean backgrounds 
caused by orbifolding.

%%%%%%%%%%%%%%%%%%%%%%%%%%%%%%%%%%%%%%%%%%%%%%%%%%%%%%%%%%%%%%%%%%%

\end{abstract}

%\vfill

\setcounter{footnote}{0}
\renewcommand{\thefootnote}{\arabic{footnote}}

\end{titlepage}

\baselineskip 18pt

\vskip2cm 
%\newpage

%%%%%%%%%%%%%%%%%%%%%%%%%%%%%%%%%%%%%%%%%%%%%%%%%%%%%%%%%%%%%%%%%%%%%%%%%%%%%%%
%%%%%%%%%%%%%%%%%%%%%%%%%%%%%%%%%%%%%%%%%%%%%%%%%%%%%%%%%%%%%%%%%%%%%%%%%%%%%%%
%%%%%%%%%%%%%%%%%%%%%%%%%%%%%%%%%%%%%%%%%%%%%%%%%%%%%%%%%%%%%%%%%%%%%%%%%%%%%%%

\section{Introduction}

One of familiar interesting features of black-hole (BH) physics is the emergence of thermodynamics. 
After making the Wick rotation, the Euclidean geometry is smoothly defined only if the imaginary time axis is 
asymptotically compactified to a circle  with a definite radius, which defines the Hawking temperature. 
Note that the asymptotic thermal circle in a typical Euclidean BH background (say, Euclidean Schwarzshild BH) is topologically trivial 
and contractible to a point at the `location of horizon', which we denote $r=r_0$ here. 
Now, if considering closed strings on such BH backgrounds, 
an interesting phenomenon would happen: closed strings wound around 
the small circle very close to the point $r=r_0$ could be tachyonic, 
at least naively,  
%now matter how low Hawking temperature is assumed 
no matter how low value of the Hawking temperature ({\em i.e.\/} large asymptotic circle) we assume.
%in the asymptotic region, 
The emergence of `thermal winding tachyons' signifies 
the Hagedorn behavior \cite{Hagedorn} 
in thermal string theories \cite{thermal string, AW}. 
One of the main purposes of this paper is to show that such an `effective Hagedorn behavior' actually arises  
%in an exactly 
%soluble background of closed superstring theory. 
from the viewpoint of exact world-sheet analysis of closed superstring theory.

We  study thermodynamical features of the superstring on near-extremal black NS5-brane background, 
which shares several physical properties with the Schwarzshild BH. Especially, their Euclidean geometries would resemble at least qualitatively. 
As first pointed  out in \cite{MS-NENS5} based on the black 5-brane solution given in \cite{HorS}, the RNS superstring 
in the near horizon region of this background is described by an exactly soluble superconformal system;
$$
\frac{SL(2, \br)}{U(1)}\mbox{-supercoset} \, \times \, SU(2)\mbox{-superWZW} \,  \times \,  \br^5.
$$
Here the $\dsp SL(2, \br)/U(1)$-supercoset model \cite{KS} is identified with 
the (supersymmetric) 2-dimensional black-hole (2DBH) \cite{2DBH}.
%after taking a suitable scaling limit. 
Utilizing this fact, we study the 1-loop thermal partition function on this background. 
The most non-trivial part of our analysis lies in the sector of Euclidean 2DBH (`cigar SCFT')  
whose asymptotic circle determines the Hawking temperature of the black NS5-background. 
We should also carefully work with the contributions of RNS world-sheet fermions with suitable boundary conditions 
as the thermal superstring theory \cite{AW}.   
With the helps of our previous works \cite{ES-NH,ncpart-orb} 
%and some extensions so that spin structures are included, 
we analyze the exact thermal partition function, aiming mainly at understanding of thermal properties of the system.

~

%%%%%%%%%%%%%%%%%%%%%%%%%%%%%%%%%%%%%%%%%%%%%%%%%%%%%%%%%%%%%%%%%%%%%%%%%%%%%%

This paper is organized as follows:

In section 2, as a preliminary, we make a brief review of near-extremal black NS5-brane system  and 
its interpretation as a superconformal system under the near-horizon limit 
\cite{MS-NENS5}.

In section 3, we evaluate the thermal partition function on this background. 
This analysis is parallel to those given in our previous works \cite{ES-NH,ncpart-orb},
but includes some extensions so that spin structures are included.  
After making the IR-regularization as given in \cite{ES-NH}, 
we can naturally decompose the obtained partition function into two parts;
$$
Z^{(\reg)}(\tau) = Z^{(\asp)}(\tau) + Z^{(\fin)} (\tau),
$$
that is, the `asymptotic part' and the `finite part'. 
The former is identified as the one proportional to $- \log \vep$, where $\vep$ is the regularization parameter.
This factor is interpreted as the divergent volume factor. 
On the other hand, the latter is finite under the $\vep\, \rightarrow \, +0$ limit.
It is interpreted as contributions from the strongly curved region near the tip of cigar.

In section 4, we shall make an analysis of spectra read off from the partition function, 
mainly focusing on the light excitations. The analysis with respect to the finite part $Z^{(\fin)} (\tau)$
is a main result of this paper. Among other things, we will observe the emergence of the `effective Hagedorn behavior' suggested above. 
Namely, $Z^{(\fin)} (\tau)$ behaves as if we were just at the Hagedorn temperature, 
although the Hawking temperature is much lower than the Hagedorn one. 
%%%%%%%%%%%%%%%%%%%%%%%%%%%%%%%%%%%%%%%%%%%%%%%%%%
We also examine the high temperature backgrounds defined by 
the $\bz_M$-orbifolding along the Euclidean time direction, which extend singularities 
at the tip. In those cases we will find out thermal behaviors with temperatures exceeding the Hagedron one. 
%%%%%%%%%%%%%%%%%%%%%%%%%%%%%%%%%%%%%%%%%%%%%%%%%

In section 5, we give a summary and some additional comments.

~

%%%%%%%%%%%%%%%%%%%%%%%%%%%%%%%%%%%%%%%%%%%%%%%%%%%%%%%%%%%%%%%%%%%%%%%%%%%%%%
%%%%%%%%%%%%%%%%%%%%%%%%%%%%%%%%%%%%%%%%%%%%%%%%%%%%%%%%%%%%%%%%%%%%%%%%%%%%%%

\section{Near Extremal Black NS5-brane Background}

The supergravity solution for the 
near extremal black NS5-branes $(\sharp\,\mbox{NS5} = k \in \bz_{>0})$ 
is written as \cite{HorS}
\begin{eqnarray}
&& ds^2 = -\left(1-\frac{r_0^2}{r^2}\right)dt^2 
+ \left(1+\frac{k\al'}{r^2}\right) 
\left(\frac{dr^2}{1-\frac{r_0^2}{r^2}}
+r^2 d\Omega_3^2\right)+ dy_5^2~,  
\label{sol black NS5} \\
&& e^{2\Phi} = g_s^2 \left(1+\frac{k\al'}{r^2}\right)~, \nonumber
\end{eqnarray}
where $r=r_0$ is the horizon, and $g_s$ denotes the string coupling at the asymptotic region 
$r\, \rightarrow\, \infty$. (The familiar extremal solution corresponds to $r_0=0$.)
It is convenient to introduce the parameter
\begin{eqnarray}
\mu \equiv \frac{r_0^2}{g_s^2 \al'},
\label{mu def}
\end{eqnarray}
which is identified  as the energy density above extremality. 
The near horizon limit is defined by taking the limit $r_0, \, g_s \, \rightarrow\, 0$ with  $\mu$ kept  finite. 
Defining new variable $\rho$ by 
$r=r_0 \cosh \left( \frac{\rho}{\sqrt{k\al'}}\right)$,
%and Wick rotation: 
%$
%t\,\rightarrow\, -i \sqrt{k\al'} \tau,
%$
we obtain the near horizon geometry \cite{MS-NENS5};
\begin{eqnarray}
&& ds^2 =   -\tanh^2 \left(\frac{\rho}{\sqrt{k\al'}} \right) d t^2 
 + d\rho^2
+ k \al' d\Omega_3^2 + dy_5^2, 
\label{NH black NS5} \\
&& e^{2\Phi} = \frac{k}{\mu \cosh^2 \left( \frac{\rho}{\sqrt{k\al'}}\right)}~.
\nonumber
\end{eqnarray}
The $(t,\rho)$-sector is identified with  
the 2-dimensional black-hole (2DBH) \cite{2DBH}, while
the $S^3$ part corresponds to the (super) $SU(2)$ WZW model together with
the implicit Kalb-Ramond field $B_{\mu\nu}$ in a familiar manner 
(with the {\em bosonic} level $k-2$). 
%%%%
We also compactify the directions parallel to the NS5s ($y$-directions) 
to a 5-torus $T^5$. 
%%%%
We have thus found that the type II string on this background
is described by the superconformal system;
\begin{equation}
\frac{SL(2, \br)_{k+2}}{U(1)} \times SU(2)_{k-2} \times T^5.
\label{SCFT NH black NS5}
\end{equation}
%%%%
The criticality condition is satisfied as 
\begin{equation}
\frac{3(k+2)}{k} + \left(\frac{3(k-2)}{k}+ \frac{3}{2}\right) 
+ \frac{3}{2}\times 5 = 15~.
\end{equation}
As mentioned in \cite{MS-NENS5}, the system is weakly coupled in the both senses of  world-sheet and space-time, 
when 
\begin{equation}
k \gg 1, \hspace{1cm} \frac{\mu}{k} \gg 1.
\label{cond weak coupling}
\end{equation}
is satisfied.
We shall assume it throughout this paper.

%%%%%%%%%%%%%%%%%%%%%%%%%%%%%%%%%%%%%%%

The Wick rotation $t\,\rightarrow\, -i \sqrt{k\al'} t_E$ converts the system 
into a thermal model:
\begin{equation}
ds^2 =   k \al' \tanh^2 \left(\frac{\rho}{\sqrt{k\al'}} \right) d t_E^2 
 + d\rho^2
+ k \al' d\Omega_3^2 + dy_5^2.
\label{Euclidean NH black NS5}
\end{equation}
%%%%%%%
which amounts to replacing 
the Lorentzian 2DBH with the Eucledean 2DBH realized as the `cigar' geometry. 
As is well-known, if requiring the smoothness of geometry, the Euclidean time $t_E$ 
need possess the periodicity $t_E \cong t_E + 2\pi$,
and hence the asymptotic radius of cigar is 
fixed to be $\sqrt{\al' k }$.
This just means that the present black-hole background has the Hawking temperature:
%\footnote
%  {From now on, we denote the Hawking temperature $\beta_{Hw}^{-1}$,
%   and the Hagedorn temperature $\beta_{Hg}^{-1}$.}
\begin{equation}
%\beta_{Hw} = 2\pi \sqrt{2N}~,
%T_{Hw} = \frac{1}{2\pi \sqrt{\al'k}}
T_{\msc{Hw}}\equiv \beta_{\msc{Hw}}^{-1}, 
\hspace{1cm} \beta_{\msc{Hw}} = 2\pi \sqrt{\al'k}.
\label{Hawking T}
\end{equation}

It is familiar that the Hagedorn temperature of free string gas is 
uniquely determined from the `effective central charge' 
$c_{\msc{eff}}$ \cite{KutS-ceff} of the transverse sector as  
\begin{equation}
T_{\msc{Hg}}\equiv \beta_{\msc{Hg}}^{-1}, \hspace{1cm}
%%%%
\beta_{\msc{Hg}} = 2\pi \sqrt{\frac{\al' c_{\msc{eff}}}{6}},
%= 2\pi \sqrt{2\al' \left(1-\frac{1}{2k}\right)}, ~~~ 
%\cQ = \sqrt{\frac{2}{k}}
\label{Hagedorn T 0}
\end{equation}
by means of the Cardy formula.
%%%%%%%%%%%%%%%%%%%%%%%%%%%%%%%%%%%%
%%%%%%%%%%%%%%%%%%%%%%%%%%%%%%%%%%%%
%\footnote
%{In order to avoid a possible confusion, we here emphasize that the Hagedorn }
%%%%%%%%%%%%%%%%%%%%%%%%%%%%%%%%%%%%
%%%%%%%%%%%%%%%%%%%%%%%%%%%%%%%%%%%% 
In the present background, this leads to 
\begin{equation}
\beta_{\msc{Hg}} 
= 2\pi \sqrt{2\al' \left(1-\frac{1}{2k}\right)}, 
\label{Hagedorn T}
\end{equation}
because of
%$c_{\msc{eff}}=12 - 24 \times \frac{\cQ^2}{8} 
%= 12 \left(1-\frac{\cQ^2}{4}\right)$.
$c_{\msc{eff}}=12 - 24 \times \frac{1}{4k} 
= 12 \left(1-\frac{1}{2k}\right)$, 
where the correction $\frac{1}{4k}$ is the mass gap that originates from
the (asymptotic) linear dilaton term\footnote
%%%%%%%%%%%%%%%%%%%%%%%%%%%%%%%%%%%%%%%
%%%%%%%%%%%%%%%%%%%%%%%%%%%%%%%%%%%%%%%
{To be more precise, 
we have to take account of the contributions from the discrete representations 
% the curved region 
%around the tip of cigar 
to determine the effective central charge $c_{\msc{eff}}$. 
%In other words, normalizable states belonging to the discrete representations 
%would alter the value of $c_{\msc{eff}}$. 
However, as was shown {\em e.g.\/} 
in \cite{ES-BH}, the identity rep. (graviton rep.) decouples 
from the $SL(2,\br)/U(1)$-sector and the conformal weights of normalizable 
discrete states are greater than $\frac{1}{2k}$. Thus,  $c_{\msc{eff}}$
is unchanged even if taking account of this contribution.  
%%%%%%%%%%%%%%%%%%
%%%%%%%%%%%%%%%%%%
In order to avoid a possible confusion, 
we also emphasize that the Hagedorn temperature $T_{\msc{Hg}}$ considered through this paper 
is defined with respect to the fundamental superstring, 
not to the so-called `Little String Theory' (LST) \cite{LST}. 
See also the second comment given at the end of section 5.
%%%%%%%%%%%%%%%%%%
%%%%%%%%%%%%%%%%%%
}.  
%%%%%%%%%%%%%%%%%%%%%%%%%%%%%%%%%%%%%%%%
%%%%%%%%%%%%%%%%%%%%%%%%%%%%%%%%%%%%%%%%
Since we assumed  a sufficiently large $k(\equiv \sharp\, \mbox{NS5})$, 
we have $T_{\msc{Hw}} \ll T_{\msc{Hg}}$.

~

%%%%%%%%%%%%%%%%%%%%%%%%%%%%%%%%%%%%%%%%%%%%%%%%%%%%%%%%%
%%%%%%%%%%%%%%%%%%%%%%%%%%%%%%%%%%%%%%%%%%%%%%%%%%%%%%%%%
%%%%%%%%%%%%%%%%%%%%%%%%%%%%%%%%%%%%%%%%%%%%%%%%%%%%%%%%%

\section{Thermal Partition Function}

In this section we evaluate the thermal torus partition function with the Hawking temperature \eqn{Hawking T}.  
To this aim, 
it is convenient to separate the contribution depending on the spin structures, 
which includes the $SL(2,\br)/U(1)$-supercoset,
8 free fermions, and also the superconformal ghosts $\beta$, $\gamma$. 
Namely, the desired partition function is written as  ($\tau\equiv \tau_1 +i \tau_2$)
\begin{eqnarray}
Z(\tau) &=& \int_{\cF} \frac{d^2 \tau}{\tau_2^2} \, \sum_{\sigma_L,\sigma_R\,:\, \msc{spin structure}}\, Z_{[\sigma_L,\sigma_R]}(\tau) \,
Z_{SU(2)}(\tau) \, Z_{T^5}(\tau) \,
 Z_{bc\msc{-gh}}(\tau),
\end{eqnarray}
where 
$Z_{[\sigma_L,\sigma_R]}(\tau)$ denotes the relevant part depending on the spin structures, while  
$Z_{SU(2)}(\tau)$ and $Z_{T^5}(\tau)$ denote 
the bosonic parts of $SU(2)$, $T^5$ sectors that 
are independent of spin structures. 
$\cF$ denotes the fundamental region as usual;
$$
\cF :=  \left\{ \tau \in \bh \, | ~  |\tau| \geq 1, ~ -\frac{1}{2} \leq  \tau_1 < \frac{1}{2}\right\},
\hspace{1cm}
\left(\bh \equiv \left\{z\in \bc\, | ~ \Im \, z >0 \right\}\right).
$$
The modular invariance of $Z(\tau)$ especially requires the `modular covariance' of the spin-structure part $Z_{[\sigma_L,\sigma_R]}(\tau)$
expressed as 
%%%%%%%%%%%%%%%%%%%%%%%%%%%%%%%%%%%%%%%%%%%%%%
\begin{eqnarray}
&& Z_{[\sigma_L,\sigma_R]} (\tau+1) = Z_{[T\cdot \sigma_L, T\cdot \sigma_R]} (\tau),
\hspace{1cm} 
Z_{[\sigma_L,\sigma_R]} \left(- \frac{1}{\tau} \right) = 
Z_{[S \cdot \sigma_L, S \cdot \sigma_R]} (\tau).
\label{modular covariance}
\end{eqnarray}
%%%%%%%%%%%%%%%%%%%%%%%%%%%%%%%%%%%%%%%%%%%%%%
Here we introduced the notations;
\begin{equation}
T\cdot \NS = \tNS, ~~~ T\cdot \tNS = \NS, ~~~ T\cdot \R = \R, ~~~ T\cdot \tR = \tR,
\label{T sigma}
\end{equation}
%%%%%%%%%
\begin{equation}
S\cdot \NS = \NS, ~~~ S\cdot \tNS = \R, ~~~ S\cdot \R = \tNS, ~~~ S\cdot \tR = \tR.    
\label{S sigma}
\end{equation}

%%%%%%%%%%%%%%%%%%%%%%%%%%%%%%%%%%%%%%%%%%%%%%%%%%%%%%%%%%%%%%%%%%%%%%%%%%%%%%%%%
%%%%%%%%%%%%%%%%%%%%%%%%%%%%%%%%%%%%%%%%%%%%%%%%%%%%%%%%%%%%%%%%%%%%%%%%%%%%%%%%%

Now, let us focus on $Z_{[\sigma_L,\sigma_R]}(\tau)$. 
According to the analysis \cite{ES-BH,ES-NH,ncpart-orb} on the $SL(2,\br)/U(1)$-supercoset, 
it is written in the form of
%%%
\begin{eqnarray}
Z^{(\reg)}_{[\sigma_L,\sigma_R]}(\tau; \vep) & = &k \sum_{m_i\in\bz}\,
\int_{\vep}^{1-\vep} ds_1 \int_0^1 ds_2 \,  \ep(\sigma_L; m_1,m_2) \ep(\sigma_R; m_1,m_2)\, 
\nn
&& ~~~ \times  
f_{[\sigma_L]}(\tau, s_1\tau+s_2)
\left[f_{[\sigma_R]}(\tau,s_1\tau+s_2)\right]^* \,
%\nn
%&& \hspace{1cm} \times
e^{-\frac{\pi k}{\tau_2}\left|(s_1+m_1)\tau+ (s_2+m_2) \right|^2} 
\label{thermal part fn 0}
\end{eqnarray}
where we set
\begin{equation}
f_{[\sigma]}(\tau,u) := \frac{\th_{[\sigma]}(\tau,u)}{\th_1(\tau,u)} \,
\left(\frac{\th_{[\sigma]}(\tau,0)}{\eta(\tau)}\right)^3,
\label{free fermion fn}
\end{equation}
with the abbreviated notation for theta functions \eqn{thsigma}.
The factor $\left(\frac{\th_{[\sigma]}}{\eta}\right)^3$ is 
just identified as the contribution from free fermions 
(and the superconformal ghosts), while $\frac{\th_{[\sigma]}(u)}{\th_1(u)}$ originates from 
the $SL(2,\br)/U(1)$-part \cite{ES-BH,ES-NH,ncpart-orb}. 
The integers $m_1$, $m_2$ are  identified with the winding numbers around the (asymptotic) thermal circle with the radius 
$\frac{\beta_{\msc{Hw}}}{2\pi} = \sqrt{\al' k}$, that is, 
$$
m_1 = \mbox{spatial winding}, \hspace{1cm} m_2 = \mbox{temporal winding}.
$$
%%%%
Moreover, we introduced the phase factor\footnote
   {Of course, the definition for $\sigma = \tR$ is just formal, since this sector does not contribute to the partition 
function. \eqn{thermal phase} here corresponds to the type IIB string.}
\begin{equation}
\ep(\sigma; m_1,m_2) \equiv 
\left\{
\begin{array}{ll}
1 & ~~~ (\sigma =\NS) \\
(-1)^{m_1+1} & ~~~ (\sigma = \tNS) \\
(-1)^{m_2+1} & ~~~ (\sigma = \R) \\
(-1)^{m_1+m_2}  & ~~~ (\sigma = \tR) 
\end{array}
\right.
\label{thermal phase}
\end{equation}
to include the correct boundary condition 
for world-sheet fermions in the thermal superstring 
\cite{AW}.  
%We here introduced the parameter of IR regularization $\vep(>0)$ as in \cite{ES-NH}. 
The IR regularization as given in \cite{ES-NH} has been made with a positive small parameter $\vep$.

%%%%%%%%%%%%%%%%%%%%%%%%%%%%%%%%%%%%%%%%%%%%%%%%%%%%%%%%%%%%%%%%%
%%%%%%%%%%%%%%%%%%%%%%%%%%%%%%%%%%%%%%%%%%%%%%%%%%%%%%%%%%%%%%%%%
%One may rewrite \eqn{thermal part fn 0} in a simpler form:
%\begin{equation}
%Z_{[\sigma_L,\sigma_R]}(\tau) = k \int_{\bc} \frac{d^2 u}{\tau_2} \, \ep(\sigma_L) \ep(\sigma_R)\, 
% f_{[\sigma_L]}(u,\tau)
%\left[f_{[\sigma_R]}(\tau,u)\right]^* \,
%e^{-\frac{\pi k}{\tau_2}\left|u\right|^2}. 
%\label{thermal part fn 1}
%\end{equation}
%where we used the notations  
%$\th_{[\sigma]}\equiv \th_3,\, \th_4, \, \th_2,\, -i \th_1$, 
%and $\ep(\sigma)=1,\,-1,\,-1,\,1$
%for $\sigma = \NS,\,\tNS,\, \R, \, \tR$ respectively. 
%%%%%%%%%%%%%%%%%%%%%%%%%%%%%%%%%%%%%%%%%%%%%%%%%%%%%%%%%%%%%%%%%%
%%%%%%%%%%%%%%%%%%%%%%%%%%%%%%%%%%%%%%%%%%%%%%%%%%%%%%%%%%%%%%%%%%%%%%%%%%%%%%%

To proceed further, we shall introduce the integer parameters 
$N$ and $K$ such that $ k= N/K$ as in \cite{ES-NH,ncpart-orb}
and {\em assume $K \in 2\bz_{>0}$}
%%%%%%%%%%%%%%%%%%%%
\footnote{We do not assume $N$ and $K$ are coprime, and thus this assumption can be always 
satisfied. Although $k$ is a positive integer here (since it is identified with the NS5 charge), 
{\em we shall not choose $K=1$.} It might sound unnatural,  
but this is a convenient assumption for our analysis 
of the `character decomposition' of the partition function 
with the {\em every} spin structure. It is found 
that all the results we give in this paper do not 
depend on the choice of $N$ and $K$ as long as the condition $K \in 2\bz_{>0}$ 
is satisfied. Of course, the simplest choice would be
$N=2k$ and $K=2$. }. 
%%%%%%%%%%%%%%%%%%%%%
One can evaluate the integrations of moduli $s_1$, $s_2$ in a way parallel to \cite{ES-NH}. 
We thus just sketch how to evaluate it with making emphasis about differences from \cite{ES-NH};
%%%%%%
\begin{itemize}
\item  
The temporal winding $m_2$ is dualized into the KK momentum $n$ by means of Poisson resummation.
%\eqn{PR}. 
When performing it, one must correctly take account of the extra phase factors depending on the spin structures. 
For instance, we simply obtain 
$$
\sum_{m_2\in \bz}\, e^{-\frac{\pi k}{\tau_2} \left\{(s_1+m_2)\tau_1 + (s_2+m_2)\right\}^2}
= \sqrt{\frac{\tau_2}{k}}\, \sum_{n\in \bz}\, e^{-\pi \tau_2 \frac{n^2}{k} + 2\pi i n \left\{ (s_1 +m_2)\tau_1 + s_2 \right\}},
$$
for the NS-NS sector. However, the extra phase \eqn{thermal phase} yields $ n \in \bz+\frac{1}{2}$ 
{\em e.g.} in the NS-R sector.

%%%%%%%%%%%%%%%%%%%%%%%%%%%%%%%%%%%%%%%%%%%%%%%%%%%%

\item To make the integral over the $U(1)$-modulus $u\equiv s_1 \tau + s_2$, 
it is convenient to utilize the identities \eqn{theta id spin str} in the factor
\eqn{free fermion fn}. For example, we obtain
\begin{equation}
f_{[\sNS]}(\tau,u)= \left(\frac{\th_{3}(\tau,0)}{\eta(\tau)}\right)^3\,
\frac{\th_{3}(\tau,0)}{i \eta(\tau)^3} \, \sum_{\nu \in \bz+\frac{1}{2}}\,
\frac{e^{2\pi i u \nu}}{1+q^{\nu}},
\label{id f NS}
\end{equation}
for the $\NS$-sector. 
Here, the powers of $q$-expansion $\nu$
take values in half-integers for the $\NS$ and $\tNS$ sectors, and  integers for the $\R$ (and $\tR$) sector. 
Note also that the contribution from  $\tR$-sector trivially vanishes. 
The $s_2$-integral imposes the constraints
\begin{equation}
\nu-\tnu = n, 
\label{constraint nu tnu}
\end{equation}
where $\nu$, $\tnu$ denote the  powers of $q$-expansions in 
the left and right movers, respectively. 
This constraint is always meaningful for any spin structure. For example, in the case of NS-R sector,
both sides of \eqn{constraint nu tnu} take values in half-integers due to 
the above remark.

%%%%%%%%%%%%%%%%%%%%%%%%%%%%%%%%%%%%%%%%%%%%%%%%%%%%%%%%

\item 
It is convenient to introduce a combined quantum number 
$v$ defined by 
\begin{equation}
v:= N m_1 - K (\nu + \tnu), 
\label{def v}
\end{equation}
and to express the partition function in terms of a summation over $\nu$, $\tnu$ and $v$
with a constraint $v+K(\nu+\nu) \in N \bz$.
%where $\nu$, $\tnu$ denote the  powers of $q$-expansions in 
%the left and right movers, respectively. 
Here $v$ is always an integer,  
since we are assuming $K \in 2\bz_{>0}$.  
%%%
We also remark that the partition function gets a phase
$
\dsp 
(-1)^{\frac{n+K(\nu+\tnu)}{N}}
$
for the $\NS$-$\tNS$, $\tNS$-$\NS$, $\R$-$\tNS$, and $\tNS$-$\R$ sectors, 
which originates from \eqn{thermal phase}.

%%%%%%%%%%%%%%%%%%%%%%%%%%%%%%%%%%%%%%%%%%%%%%%%%%%%%%%

\item 
We make use of the following identity 
to carry out the $s_1$-integral as in \cite{ES-NH} (see also \cite{HPT});
\begin{equation}
\hspace{-1cm}
\sqrt{k\tau_2} \int_{\vep}^{1-\vep} ds_1 \, e^{-\pi \tau_2 \frac{N}{K} s_1^2 - 2\pi \tau_2 s_1 \frac{v}{K}}
= \frac{1}{2\pi i} \int_{\br-i0} dp\, \frac{e^{-\pi \tau_2 \frac{p^2}{NK}}}{p-iv} \, 
\left\{
e^{-2\pi i \vep \frac{\tau_2}{K}(p-iv)}- e^{-2\pi i (1-\vep) \frac{\tau_2}{K}(p-iv)}
\right\}.
\end{equation}

\end{itemize}

~

%%%%%%%%%%%%%%%%%%%%%%%%%%%%%%%%%%%%%%%%%%%%%%%%%%%%%%%%%%%%%%%%%%%%%%%%%%%%%%%%%%%%%%%
%%%%%%%%%%%%%%%%%%%%%%%%%%%%%%%%%%%%%%%%%%%%%%%%%%%%%%%%%%%%%%%%%%%%%%%%%%%%%%%%%%%%%%%
Combining all the pieces, we finally obtain 
\begin{eqnarray}
&& 
\hspace{-5mm} 
Z^{(\reg)}_{[\sigma_L,\sigma_R]}(\tau; \vep) = 
\sum_{v\in\bz} \sum_{\stackrel{\nu \in \bz+\frac{s(\sigma_L)-1}{2}, 
\, \tnu \in \bz+\frac{s(\sigma_R)-1}{2}}{v+K(\nu+\tnu) \in N\bz}} 
\, \bep(\sigma_L; v,\nu,\tnu) \bep(\sigma_R; v,\nu,\tnu)
\nn
%%%%%%%
&&
\hspace{5mm}
\times
\frac{1}{2\pi i} \, \left[
\int_{\br-i0} dp \, q^{\nu} 
\overline{q^{\tnu}} (-1)^{t(\sigma_L) + t(\sigma_R)}
e^{-\vep'(v+ip)} -  \int_{\br+i(N-0)} dp\, e^{\vep'(v+ip)}
\right]
\nn
&& 
\hspace{5mm}
\times \frac{e^{-\pi \tau_2 \frac{p^2+v^2}{NK}}}{p-iv}
\, \frac{q^{\frac{K}{N}\nu^2 + \frac{v}{N}\nu}}{1+(-1)^{t(\sigma_L)} q^{\nu}}  \,
%\left\lb
\overline{
\frac{q^{\frac{K}{N}\tnu^2 + \frac{v}{N}\tnu}}{1+(-1)^{t(\sigma_R)}q^{\tnu}} 
}
%\right\rb^*
\, \left(\frac{\th_{[\sigma_L]}}{\eta}\right)^4 
\overline{ \left(\frac{\th_{[\sigma_R]}}{\eta}\right)^4}
\left|\frac{1}{\eta^2}\right|^2,
\label{Z general}
\end{eqnarray}
where we set $\vep' := 2\pi \frac{\tau_2}{K} \vep$, and 
the phase factor $\bep(\sigma; v,\nu,\tnu)$ is defined by 
\begin{equation}
\bep(\sigma; v,\nu,\tnu) := 
\left\{
\begin{array}{ll}
1 & ~~ (\sigma = \NS) \\
(-1)^{\frac{v+K(\nu+\tnu)}{N}+1} & ~~ (\sigma = \tNS) \\
-1 & ~~ (\sigma =\R) \\
(-1)^{\frac{v+K(\nu+\tnu)}{N}} & ~~ (\sigma = \tR)
\end{array}
\right.
\label{bep}
\end{equation}
We also introduced the notation
\begin{equation}
s(\sigma) := \left\{
\begin{array}{ll}
0 & ~~~ \sigma=\NS, ~ \tNS, \\
1 & ~~~ \sigma = \R, ~ \tR
\end{array}
\right.
\hspace{1.5cm}
t(\sigma) := \left\{
\begin{array}{ll}
0 & ~~~ \sigma=\NS, ~ \R, \\
1 & ~~~ \sigma = \tNS, ~ \tR
\end{array}
\right.
\end{equation}

%%%%%%%%%%%%%%%%%%%%%%%%%%%%%%%%%%%%%%%%%%%%%%%%%%%%%%%%%%%%%%%%%%%%%%%%%%%%%%%%

The regularized partition function $Z^{(\reg)}_{[\sigma_L,\sigma_R]}(\tau; \vep)$ correctly behaves under the modular T-transformation;
\begin{equation}
Z^{(\reg)}_{[\sigma_L,\sigma_R]}(\tau+1; \vep) = Z^{(\reg)}_{[T\cdot \sigma_L, T\cdot \sigma_R]}(\tau; \vep),
\label{Zreg T}
\end{equation}
where we used the notation \eqn{T sigma}.
%\begin{equation}
%T\cdot \NS = \tNS, ~~~ T\cdot \tNS = \NS, ~~~ T\cdot \R = \R, ~~~ T\cdot \tR = \tR.    
%\label{T sigma}
%\end{equation}
On the other hand, the modular covariance for the S-transformation is `weakly' violated;
\begin{equation}
Z^{(\reg)}_{[\sigma_L,\sigma_R]}\left(-\frac{1}{\tau}; \vep\right) 
= Z^{(\reg)}_{[S\cdot \sigma_L, S\cdot \sigma_R]}(\tau; \vep) + O(\vep, \vep \log \vep),
\label{Zreg S}
\end{equation}
with the notation \eqn{S sigma}.
%\begin{equation}
%S\cdot \NS = \NS, ~~~ S\cdot \tNS = \R, ~~~ S\cdot \R = \tNS, ~~~ S\cdot \tR = \tR.    
%\label{S sigma}
%\end{equation}
The S-transformation relation \eqn{Zreg S} is most easily shown by the definition \eqn{thermal part fn 0} itself. In fact, 
if rewriting \eqn{thermal part fn 0} as 
$$
Z^{(\reg)}_{[\sigma_L,\sigma_R]}(\tau; \vep) \equiv \int_{\vep}^{1-\vep} ds_1 \int_0^1 ds_2\, F_{[\sigma_L,\sigma_R]}(s_1, s_2; \tau),
$$
and by using the modular property
$$
F_{[\sigma_L,\sigma_R]}\left(s_1, s_2; -\frac{1}{\tau}\right) = F_{[S\cdot \sigma_L, S\cdot \sigma_R]}(s_2, -s_1; \tau) \equiv 
F_{[S\cdot \sigma_L, S\cdot \sigma_R]}(s_2, 1-s_1; \tau),
$$ 
we obtain
\begin{eqnarray*}
&&
\hspace{-5mm}
Z^{(\reg)}_{[\sigma_L,\sigma_R]}\left(-\frac{1}{\tau}; \vep\right)
 -  Z^{\msc{(reg)}}_{[S\cdot \sigma_L, S\cdot \sigma_R]}(\tau; \vep)
\nn
&& \hspace{1cm}
= \left\lb 
 \int_0^{\vep} ds_1 \int_{\vep}^{1-\vep} ds_2
+ \int_{1-\vep}^{1} ds_1 \int_{\vep}^{1-\vep} ds_2
- \int_{\vep}^{1-\vep} ds_1 \int_0^{\vep} ds_2
- \int_{\vep}^{1-\vep} ds_1 \int_{1-\vep}^1 ds_2
\right\rb
\nn
&& \hspace{6cm}
\times
F_{[S\cdot \sigma_L,S\cdot\sigma_R]}(s_1, s_2; \tau).
\end{eqnarray*}
This integral over the moduli $s_i$ at most behaves as $\sim \vep \log \vep$ under $\vep \, \rightarrow \, +0$.

~

%%%%%%%%%%%%%%%%%%%%%%%%%%%%%%%%%%%%%%%%%%%%%%%%%%%%%%%%%%%%%%
%%%%%%%%%%%%%%%%%%%%%%%%%%%%%%%%%%%%%%%%%%%%%%%%%%%%%%%%%%%%%%
%%%%%%%%%%%%%%%%%%%%%%%%%%%%%%%%%%%%%%%%%%%%%%%%%%%%%%%%%%%%%%

\section{Analysis of Spectra}

In this section we study the spectra read off from the partition function, 
mainly focusing on the light excitations.
We shall start with decomposing the partition function \eqn{Z general}.

~

%%%%%%%%%%%%%%%%%%%%%%%%%%%%%%%%%%%%%%%%%%%%%%%%%%%%%%%%%%%%%%%

\subsection{Decomposition of Partition Function}

As performed in \cite{HPT,ES-BH,ES-NH}, we make use of the manipulation of contour deformation;
$$
\int_{\br+i(N-0)} dp \, [\cdots] \, = \, \int_{\br-i0} dp \, [\cdots] \,- 2\pi i \left[ \mbox{residues of poles in } ~ 0\leq \mbox{Im}\, p < N \right].
$$
in order to decompose the partition function \eqn{Z general}.  
This yields 
\begin{eqnarray}
&& Z^{(\reg)}_{[\sigma_L,\sigma_R]}(\tau; \vep) = Z^{(\dis)}_{0, [\sigma_L,\sigma_R]}(\tau) 
+ Z^{(\con)}_{0, [\sigma_L,\sigma_R]} (\tau; \vep),
\label{Z reg decomp}
\end{eqnarray}
where the first term is the pole part that is free from the regularization parameter $\vep$.
%%%%%%%%%%%%%%%%%%%%%%%%%%%%%%%%%%%%%%%%%%%%%%%%%%%%%%%%%%%%%%%%
One can explicitly write it as 
\begin{eqnarray}
\hspace{-1cm} 
Z^{(\dis)}_{0, [\sigma_L,\sigma_R]}(\tau) &=& \sum_{v=0}^{N-1} \, 
\sum_{\stackrel{\nu \in \bz+\frac{s(\sigma_L)-1}{2}, 
\, \tnu \in \bz+\frac{s(\sigma_R)-1}{2}}{v+K(\nu+\tnu) \in N\bz}}\,
 \bep(\sigma_L; v,\nu,\tnu) \bep(\sigma_R; v,\nu,\tnu) 
\nn
&& 
\hspace{5mm}
\times  
\frac{q^{\frac{K}{N}\nu^2 + \frac{v}{N}\nu}}{1+(-1)^{t(\sigma_L)} q^{\nu}}  \,
%\left\lb
\overline{
\frac{q^{\frac{K}{N}\tnu^2 + \frac{v}{N}\tnu}}{1+(-1)^{t(\sigma_R)}q^{\tnu}} 
}
%\right\rb^*
\, \left(\frac{\th_{[\sigma_L]}}{\eta}\right)^4 
\overline{ \left(\frac{\th_{[\sigma_R]}}{\eta}\right)^4}
\left|\frac{1}{\eta^2}\right|^2
\nn
%%%
&=& 
\sum_{v=0}^{N-1} \, 
\sum_{\stackrel{a \in \bz_N+\frac{s(\sigma_L)-1}{2}, 
\, \ta \in \bz_N+\frac{s(\sigma_R)-1}{2}}{v+K(a+\ta) \in N\bz}}\,
\bep(\sigma_L; v,a,\ta) \bep(\sigma_R; v,a,\ta) 
\nn
&&
\hspace{5mm}
\times
 \chids{\sigma_L}(v, a;\tau) \, \overline{\chids{\sigma_R}(v, \ta;\tau)} 
 \, \left(\frac{\th_{[\sigma_L]}}{\eta}\right)^3
\overline{ \left(\frac{\th_{[\sigma_R]}}{\eta}\right)^3}.
\label{Z dis 0 general}
\end{eqnarray}
%%%%%%%%%%%%%%%%%%%%%%%%%%%%%%%%%%%%%%%%%%%%%%%%%%%%%%%%%%%%%%%%%%%%
Here $\chids{\sigma}(v, a;\tau)$ denotes the extended discrete character with the spin structure $\sigma$
\eqn{chid}. 
%%%%%%%%%%%%%%%%%%%%%%%%%%%%%
%Despite the `natural form' as a torus partition function, 
%\eqn{Z dis 0 general} does not show the simple modular covariance. 
%From this viewpoint, it would be rather natural to adopt another decomposition 
%\eqn{ES-NH};
%\begin{eqnarray}
%&& Z^{\msc{(reg)}}_{[\sigma_L,\sigma_R]}(\tau; \ep) = Z^{(\dis)}_{0, [\sigma_L,\sigma_R]}(\tau) 
%+ Z^{(\con)}_{0, [\sigma_L,\sigma_R]} (\tau; \ep),
%\label{Z reg decomp}
%\end{eqnarray}
%%%%%%%%%%%%%%%%%%%%%%%%%%%%%%%%%%%%%%%%%%%%%%
Also, the second term is written as 
\begin{eqnarray}
&& \hspace{-1cm} 
Z^{\msc{(\con)}}_{0, [\sigma_L,\sigma_R]}(\tau; \vep) = 
\sum_{v\in\bz} \sum_{\stackrel{\nu \in \bz+\frac{s(\sigma_L)-1}{2}, 
\, \tnu \in \bz+\frac{s(\sigma_R)-1}{2}}{v+K(\nu+\tnu) \in N\bz}} 
\, \bep(\sigma_L; v,\nu,\tnu) \bep(\sigma_R; v,\nu,\tnu)
\nn
%%%%%%%
&&
\hspace{1.5cm}
\times
\frac{1}{2\pi i} \, 
\int_{\br-i0} dp \, \frac{e^{-\pi \tau_2 \frac{p^2}{NK}}}{p-iv} \, 
q^{\frac{K}{N}\left(\nu+ \frac{v}{2K}\right)^2}\,
\overline{
q^{\frac{K}{N}\left(\tnu+ \frac{v}{2K}\right)^2}\,
}
\,\left(\frac{\th_{[\sigma_L]}}{\eta}\right)^4 
\overline{ \left(\frac{\th_{[\sigma_R]}}{\eta}\right)^4}
\left|\frac{1}{\eta^2}\right|^2,
\nn
&& 
\hspace{1.5cm}
\times 
\left\lb
 \frac{1}{1+(-1)^{t(\sigma_L)}q^{-\nu}} e^{-\vep'(v+ip)}
- \overline{ 
\frac{1}{1+(-1)^{t(\sigma_R)}q^{\tnu}}
}
e^{\vep'(v+ip)}
\right.
\nn
&& \hspace{6cm} 
\left.
+ \frac{ e^{\vep'(v+ip)} - e^{-\vep'(v+ip)}}
{\left\{ 1+(-1)^{t(\sigma_L)}q^{-\nu} \right\}\left\{1+(-1)^{t(\sigma_R)}\overline{q^{\tnu}} \right\}}
\right\rb .
%%%%
%\times 
%\left\lb
% \frac{(-1)^{t(\sigma_L)}q^{\nu}}{1+(-1)^{t(\sigma_L)}q^{\nu}} e^{-\vep(v+ip)}
%- \overline{ 
%\frac{1}{1+(-1)^{t(\sigma_R)}q^{\tnu}}
%}
%e^{\vep(v+ip)}
%\right.
%\nn
%&& \hspace{4cm} 
%\left.
%+ \frac{ (-1)^{t(\sigma_L)}q^{\nu} \left\{e^{\vep(v+ip)} - e^{-\vep(v+ip)}\right\}}
%{\left\{ 1+(-1)^{t(\sigma_L)}q^{\nu} %\right\}\left\{1+(-1)^{t(\sigma_R)}\overline{q^{\tnu}} \right\}}
%\right\rb
\label{Z con 0 general}
\end{eqnarray}
This part depends on the regularization parameter $\vep$ and shows a logarithmically divergence under the $\vep\, \rightarrow \, +0$ limit.

%%%%%%%%%%%%%%%%%%%%%%%%%%%%%%%%%%%%%%%%%%%%%%%%%%%%%%%%%%%%%%%%%%%%%%%%%%%%%%%%%%%

~

\noindent
{\bf Note: } 
Despite the `natural form' 
%as a torus partition function
from the viewpoint of superconformal algebra, 
\eqn{Z dis 0 general} does not show the simple modular covariance. 
From this reason it would be rather natural to adopt another decomposition as is discussed in 
\cite{ES-NH};
\begin{equation}
Z^{(\reg)}_{[\sigma_L,\sigma_R]}(\tau; \vep) = Z^{(\dis)}_{[\sigma_L,\sigma_R]}(\tau) 
+ Z^{(\con)}_{[\sigma_L,\sigma_R]} (\tau; \vep),
\label{Z reg decomp 2}
\end{equation}
where $Z^{(\dis)}_{[\sigma_L,\sigma_R]}(\tau)$ is defined by replacing 
$\chids{\sigma}(v, a;\tau)$ with its modular completion $\hchids{\sigma}(v, a;\tau)$
\eqn{hchid} in the R.H.S of \eqn{Z dis 0 general};
%%%%%%%%%%%%%%%%%%%%%%%%%%%%%%%%%%%%%%%%%%%%%%%%%%%%
\begin{eqnarray}
Z^{(\dis)}_{[\sigma_L,\sigma_R]} (\tau)
&=& 
\sum_{v=0}^{N-1} \, 
\sum_{\stackrel{a \in \bz_N+\frac{s(\sigma_L)-1}{2}, 
\, \ta \in \bz_N+\frac{s(\sigma_R)-1}{2}}{v+K(a+\ta) \in N\bz}}\,
\bep(\sigma_L; v,a,\ta) \bep(\sigma_R; v,a,\ta) 
\nn
&&
\hspace{5mm}
\times
 \hchids{\sigma_L}(v, a;\tau) \, \overline{\hchids{\sigma_R}(v, \ta;\tau)} 
 \, \left(\frac{\th_{[\sigma_L]}}{\eta}\right)^3
\overline{ \left(\frac{\th_{[\sigma_R]}}{\eta}\right)^3},
\label{Z dis general}
\end{eqnarray}
%%%%%%%%%%%%%%%%%%%%%%%%%%%%%%%%%%%%%%%%%%%%%%%%%%%%
and precisely satisfies the modular covariance relations;
\begin{eqnarray}
&&
Z^{(\dis)}_{[\sigma_L,\sigma_R]}\left(\tau+1\right) = Z^{(\dis)}_{[T\cdot \sigma_L, T\cdot \sigma_R]} (\tau), \hspace{1cm}
 Z^{(\dis)}_{[\sigma_L,\sigma_R]}\left(-\frac{1}{\tau}\right) = Z^{(\dis)}_{[S\cdot \sigma_L, S\cdot \sigma_R]} (\tau).
\end{eqnarray}
It is also worthwhile to remark that the remaining function 
$Z^{(\con)}_{[\sigma_L,\sigma_R]} (\tau; \vep)$ is expressible 
as a sesquilinear form  of  only the extended continuous characters $\chics{\sigma}(p,m;\tau)$ \eqn{chic} contrary
 to $Z^{(\con)}_{0,[\sigma_L,\sigma_R]} (\tau; \vep)$ \eqn{Z con 0 general}, 
although we shall omit its explicit form here. 
However, we shall make use of the `old decomposition' \eqn{Z reg decomp} 
%rather than \eqn{Z reg decomp 2} 
for the time being, since it seems relatively easier to analyze the mass spectra of excitations based on it.

~

%%%%%%%%%%%%%%%%%%%%%%%%%%%%%%%%%%%%%%%%%%%%%%%%%%%%%%%%%%%%%%%
%%%%%%%%%%%%%%%%%%%%%%%%%%%%%%%%%%%%%%%%%%%%%%%%%%%%%%%%%%%%%%%

%\noindent
%\underline{\bf asymptotic part :}

\subsection{Asymptotic Part}

%%%%%%%%%%%%%%%%%%%%%%%%%%%%%%%
% asymptotic part
%%%%%%%%%%%%%%%%%%%%%%%%%%%%%%
We first extract the `asymptotic part' of partition function  
that shows a logarithmic divergence. Namely, we set 
\begin{eqnarray}
\hZ^{(\asp)}_{[\sigma_L,\sigma_R]}(\tau)
& :=  & - \lim_{\vep\,\rightarrow\,+0}\, 
\left[ \vep \frac{\del}{\del \vep} \, 
Z^{(\reg)}_{[\sigma_L,\sigma_R]}(\tau;\vep)\right]
\nn
&\equiv & 
- \lim_{\vep\,\rightarrow\,+0}\, 
\left[ \vep \frac{\del}{\del \vep} \, 
Z^{(\con)}_{0, [\sigma_L,\sigma_R]}(\tau;\vep)\right],
\label{def hasp}
\end{eqnarray}
and 
\begin{eqnarray}
Z^{(\reg)}_{[\sigma_L,\sigma_R]}(\tau;\vep) & = & 
Z^{(\asp)}_{[\sigma_L,\sigma_R]} (\tau;\vep) + Z^{(\fin)}_{[\sigma_L,\sigma_R]} (\tau)
+ O(\vep, \vep \log \vep)
\nn
&\equiv & 
- \log \vep \,  \hZ^{(\asp)}_{[\sigma_L,\sigma_R]}(\tau) +  Z^{(\fin)}_{[\sigma_L,\sigma_R]} (\tau) +
O(\vep, \vep \log \vep).
\label{decomp asp fin}
\end{eqnarray}
%Since $- \log \vep$ is identified with the volume factor of target space, one can expect that 
Here 
the `finite part' $Z^{(\fin)}_{[\sigma_L,\sigma_R]} (\tau)$ 
is uniquely determined in this decomposition by requiring 
the independence of the regularization parameter $\vep$. 
We will later examine this part.

%%%%%%

Based on  \eqn{Z con 0 general} and \eqn{def hasp}
it is easy to evaluate the explicit form of $\hZ^{(\asp)}_{[\sigma_L,\sigma_R]}(\tau)$;
\begin{eqnarray}
\hspace{-5mm}
\hZ^{(\asp)}_{[\sigma_L,\sigma_R]}(\tau)
&=& \frac{1}{\pi} \int_{-\infty}^{\infty} dp\,
\sum_{\stackrel{w\in\bz}
{n\in \bz+ \frac{s(\sigma_L)+s(\sigma_R)}{2}}}
\,
\ep(\sigma_L; w, 0) \ep(\sigma_R; w,0)
\nn
&& \hspace{5mm}
\times 
q^{\frac{p^2}{4NK}+ \frac{(Nw+Kn)^2}{4NK}}\,
\overline{q^{\frac{p^2}{4NK}+ \frac{(Nw-Kn)^2}{4NK}}}\,
\left(\frac{\th_{[\sigma_L]}}{\eta}\right)^4 \, 
\overline{\left(\frac{\th_{[\sigma_R]}}{\eta}\right)^4 }\,
\left|\frac{1}{\eta^2}\right|^2
\nn
%%%%%%
&=& \frac{2}{\pi} \int_{0}^{\infty} dp\,
\sum_{\stackrel{w_0\in\bz_{2K}}
{n_0\in \bz_N+ \frac{s(\sigma_L)+s(\sigma_R)}{2}}}
\,
\ep(\sigma_L; w_0, 0) \ep(\sigma_R; w_0,0)
\nn
&& \hspace{5mm}
\times \chics{\sigma_L}(p, Nw_0+Kn_0;\tau,0) \,
\overline{\chics{\sigma_R}(p, Nw_0-Kn_0;\tau,0)} \,
\left(\frac{\th_{[\sigma_L]}}{\eta}\right)^3 \, 
\overline{\left(\frac{\th_{[\sigma_R]}}{\eta}\right)^3 },
\nn
&& 
\label{hZ asp}
\end{eqnarray}
where $\chics{\sigma}(p,m;\tau)$ denotes the extended continuous character 
\eqn{chic} written explicitly as 
\begin{equation}
\chics{\sigma}(p,m;\tau) \equiv q^{\frac{p^2}{4NK}} \Th{m}{NK}(\tau,0)\, \frac{\th_{[\sigma]}(\tau,0)}{\eta(\tau)^3},
\label{chic main text}
\end{equation}
in terms of the theta function. 
%%%%%%%%%%
The thermal winding number $m_1$ has been rewritten as $w$ 
in the first line of \eqn{hZ asp}.
%%%%%%%%%%%%%%%%%
This partition function essentially coincides with the one studied 
in \cite{KutSah}.
%%%%%%%%%%%%%%%%%

As is directly confirmed from \eqn{hZ asp}, 
$\hZ^{(\asp)}_{[\sigma_L,\sigma_R]}(\tau)$ correctly behaves under modular transformations:
\begin{eqnarray}
&& \hZ^{(\asp)}_{[\sigma_L,\sigma_R]} (\tau+1) = \hZ^{(\asp)}_{[T\cdot \sigma_L, T\cdot \sigma_R]} (\tau),
\hspace{1cm} 
\hZ^{(\asp)}_{[\sigma_L,\sigma_R]} \left(- \frac{1}{\tau} \right) = 
\hZ^{(\asp)}_{[S \cdot \sigma_L, S \cdot \sigma_R]} (\tau).
\label{hZ asp modular}
\end{eqnarray}

The physical interpretation of the asymptotic part is obvious:
this sector corresponds to strings freely propagating in the region 
away from the NS5-branes. 
After combining other sectors independent of the spin structures, $\hZ^{(\asp)}_{[\sigma_L,\sigma_R]}(\tau)$ is identified with 
the thermal partition function of the type IIB superstring 
%with the correct boundary conditions of world-sheet fermions 
\cite{AW} on the background 
$$
S^1_{\beta_{\msc{Hw}}} \times \br_{\phi} \times SU(2)_k \times T^5,
$$
where $S^1_{\beta_{\msc{Hw}}}$ denotes the thermal circle with the inverse Hawking temperature \eqn{Hawking T}, 
and $\br_{\phi}$ expresses a linear dilaton background with the dilaton gradient $\cQ \equiv \sqrt{\frac{2}{k}}$. 
The divergent factor $-\log \vep$ originates from the infinite volume of asymptotic region. 
This sector does not give rise to a thermal instability, 
since the Hawking temperature \eqn{Hawking T} is lower than 
the Hagedorn one \eqn{Hagedorn T}. 
(Recall that we assumed a sufficiently large value of $k$.)
%%%

%Note also that the boundary condition 
%of world-sheet fermions is the correct one essentially given in \cite{AW}.
%This fact means that 
%%%%%%%%%%%%%%%%%%%%%%%%%%%%%%%%%%%%%%%%%%%%%%%%%%%%%%%%%%%%%%%%%%%

The asymptotic sector is correctly GSO projected and preserves the space-time SUSY, 
when going back to the physical background with Lorentzian signature. 
In other words, the sector with no thermal winding is GSO projected, 
%leading to a vanishing cosmological constant, 
as is easily confirmed by observing the $w=0$ terms in \eqn{hZ asp}.

%%%%%%%%%%%%%%%%%%%%%%%%%%%%%%%%%%%%%%%%%%%%%%%%%%%%%%%%%%%%%%%%%%%%%%%%

~

The spectrum of light excitations in the asymptotic sector is summarized as follows: 
%%%
\begin{description}
\item[(i) no winding spectrum:]

~

All the states with no thermal winding are massive: the minimal conformal weight is equal  $h=\tilde{h} = \frac{1}{2}+\frac{1}{4k}$, 
which is determined by the GSO projection and the linear dilaton.

~

\item[(ii) thermal winding spectrum:]

~

The lightest winding states are the NS-NS states with $w=\pm 1$, whose conformal weights are equal $h=\tilde{h} = \frac{k}{4}$.
This fact is consistent with the value of Hawking temperature \eqn{Hawking T}.

\end{description}

%%%%%%%%%%%%%%%%%%%%%%%%%%%%%%%%%%%%%%%%%%%%%%%%%%%%%%%%%%%%%%%%%%%%%%%%
%%%%%%%%%%%%%%%%%%%%%%%%%%%%%%%%%%%%%%%%%%%%%%%%%%%%%%%%%%%%%%%%%%%%%%%%
%%%%%%%%%%%%%%%%%%%%%%%%%%%%%%%%%%%%%%%%%%%%%%%%%%%%%%%%%%%%%%%%%%%%%%%%

~

\subsection{Finite Part and the Effective Hagedorn Behavior}

Let us next examine the finite part $Z^{(\fin)}_{[\sigma_L,\sigma_R]} (\tau)$, which shows more intriguing features. 
By the definition \eqn{decomp asp fin} one may explicitly write 
\begin{equation}
Z^{(\fin)}_{[\sigma_L,\sigma_R]} (\tau) = \lim_{\vep\,\rightarrow\, +0} \, \left[ 
1 - (\log \vep) \vep \frac{\del}{\del \vep} 
\right] \, Z^{(\reg)}_{[\sigma_L,\sigma_R]}(\tau;\vep). 
\end{equation}

We first note that 
$Z^{(\fin)}_{[\sigma_L,\sigma_R]} (\tau)$ is also modular covariant;
\begin{eqnarray}
&& Z^{(\fin)}_{[\sigma_L,\sigma_R]} (\tau+1) = Z^{(\fin)}_{[T\cdot \sigma_L, T\cdot \sigma_R]} (\tau),
\hspace{1cm} 
Z^{(\fin)}_{[\sigma_L,\sigma_R]} \left(- \frac{1}{\tau} \right) = 
Z^{(\fin)}_{[S \cdot \sigma_L, S \cdot \sigma_R]} (\tau),
\label{Z fin modular}
\end{eqnarray}
even though it would appear hard to check it directly. 
This fact results from the modular properties \eqn{Zreg T}, \eqn{Zreg S} as well as \eqn{hZ asp modular}.

This sector is physically interpreted as the contribution from  strings localized near the NS5-branes, 
which are far from free. 
We now investigate the IR behavior and the spectra of light excitations
read from the partition function $Z^{(\fin)}_{[\sigma_L,\sigma_R]} (\tau)$.

~

%%%%%%%%%%%%%%%%%%%%%%%%%%%%%%%%%%%%%%%%%%%%%%%%%%%%%%%%%%%%%%%%%%%%%%%%%%%%%%

\noindent
{\bf (i) no winding spectrumF}

Thanks to the good modular behavior \eqn{Z fin modular}, one can reinterpret 
the thermal partition function $\sum_{\sigma_L,\sigma_R}\, Z^{(\fin)}_{[\sigma_L,\sigma_R]} (\tau) $ as a free energy 
of `superstring gas' localized near the NS5-branes, after combining the $SU(2)$ and $T^5$ sectors.  
This is simply achieved by dropping off the thermal winding number $m_1$ in \eqn{Z general}, 
%the spatial winding numbers around the thermal circle, which is identified with $m_1$ in \eqn{Z general}, 
and by replacing the fundamental region $\cF$ of torus modulus $\tau$ 
with the strip region \cite{Polchinski}
\begin{equation}
\cS := \left\{\tau \in \bh \, | ~ -\frac{1}{2} \leq \tau_1 < \frac{1}{2} \right\}.
\label{strip}
\end{equation}
%%%%%%%%%%%%%
We are especially interested in the IR-behavior of the free energy of localized 
components, 
dominated by light excitations with no winding $m_1=0$. 
%belonging to $Z^{(\fin)}_{[\sigma_L,\sigma_R]} (\tau) $.
%$Z^{(\fin)}_{[\sigma_L,\sigma_R]} (\tau) $ 
Since we have contributions both from $Z^{(\dis)}_{0, [\sigma_L,\sigma_R]} (\tau) $
\eqn{Z dis 0 general} and $Z^{(\con)}_{0, [\sigma_L,\sigma_R]} (\tau) $ \eqn{Z con 0 general}, 
we shall separately analyze each part:
%%%%%%%%
%%%%%%%%
%%%%%%%%
\begin{itemize}
\item {\bf $Z^{(\dis)}_{0, [\sigma_L,\sigma_R]} (\tau) $-part :}  ~~

We first point out that this sector is not supersymmetric, 
even if working on the non-thermal background.
Of course, it is not surprising since we are working with  
the non-BPS NS5-background. 
%
% in a contrast with 
%the asymptotic sector. 
%
In other words, the GSO projection does {\em not\/}  necessarily 
act in the usual form
{\em even for the states with no thermal winding\/}
%contrary to  the asymptotic sector, 
in a sharp contrast with the thermal superstring in flat backgrounds \cite{AW}.
%despite of a formal resemblance between \eqn{Z dis 0 general} and \eqn{hZ asp}.
%%%%
In an algebraic footing, this originates from the different behaviors 
of $\chics{\sigma}$ and $\chids{\sigma}$ under changing the spin structures.
The latter contains extra factors such as
$
\frac{1}{1+(-1)^{t(\sigma)}q^{\nu}} 
%\hspace{1cm} (\mbox{$\sigma$~:~ spin structure}),
$
(see \eqn{chid}), 
%depending non-trivially on the spin structure $\sigma$, and 
which would give rise to a relative sign difference of vacuum states 
between the $\NS$ and $\tNS$ sectors.

%Even after restricting to $m_1=0$, which amounts to 
%$$v+ K(\nu+\tnu) =0,$$ 
%in the summation of \eqn{Z dis 0 general}
%(see \eqn{def v}), 

%%%%%%%%%%%%%%%%%%%%%%%%%%%%%%%%%%%%%%%%%%%%%%%%%%%%%%%%%%%%%%%

Now, let us explore the lightest states. We treat each spin structure separately:
%%%
\begin{list}{}{}
\item {\bf [$\NS$-$\NS$ sector ] }

We note that setting $m_1=0$ amounts to imposing 
$v+ K(\nu+\tnu) =0,$ (see \eqn{def v}).
Therefore, we have to look for leading terms in \eqn{Z dis 0 general}
which lie in the spectral flow orbits  
(in the discrete extended character $\chids{\sNS}$, in other words) 
with the constraint:
\begin{equation}
%v+ K(\nu+\tnu) =0,
v+K(a+\ta)=0, \hspace{1cm} v=0,\ldots, N-1, ~~ a,\ta \in \frac{1}{2}+\bz_N.
\label{cond NSNS}
\end{equation}
imposed. 
The lightest excitation is found to be the vacuum state of 
discrete character $\chids{\sNS}$ with 
\begin{equation}
v=K, ~~~ a=\ta \equiv -\frac{1}{2} ~ (\mod\, N) ,
\label{no winding massless}
\end{equation} 
which corresponds to an anti-chiral primary of the $SL(2,\br)/U(1)$-sector
in both of left and right movers.
This state is  `wrong GSO' projected ($\NS$ and $\tNS$ appear 
with the same sign. See the above comment.),
and possesses conformal weights
$
h=\tilde{h} = \frac{1}{2}.
$
This means that it is a massless state after combining it with the identity states 
of $SU(2)$ and $T^5$ sectors.

%%%%%%%%%%%%%%%%%%%%%%%%%%%%%%%%%%%%%%%%%%%%%%%%%%%%%%%%%%%%%%%%%%%%%%%%

Other candidates of light excitations would be 
$a=\frac{1}{2}$, $\ta=-\frac{1}{2}$ or $a= -\frac{1}{2}$, $\ta= \frac{1}{2}$
and $v=0$. However, taking account of the correct/wrong GSO projections
({\em i.e.\/} the relative sign of vacuum states in the $\NS$ 
and $\tNS$ representations), one can find out that both has
$
h=\tilde{h} = \frac{1}{2} + \frac{1}{2k}.
$
They are hence massive states.

%%%%%%%%%%%%%%%%%%%%%%%%%%%%%%%%%%%%%%%%%%%%%%%%%%%%%%%%%%%%%%%%%%%%%%%%

~

\item {\bf [$\R$-$\R$ sector  ] }

It is obvious that the lightest state corresponds to a Ramond vacuum 
in $SL(2,\br)/U(1)$-sector. 
There is a unique RR-vacuum in the spectral flow orbit 
of $v=0$ and $a=\ta=0 \in \bz_N$ that satisfies the constraints:
\begin{equation}
%v+ K(\nu+\tnu) =0,
v+K(a+\ta)=0, \hspace{1cm} v=0,\ldots, N-1, ~~ a,\ta \in \bz_N.
\label{cond RR}
\end{equation}
Combining it with the Ramond vacua of free fermions, 
we obtain the states with conformal weight 
$
h=\tilde{h} = \frac{1}{2} + \frac{1}{4k} > \frac{1}{2},
$
which are massive excitations.

~

%%%%%%%%%%%%%%%%%%%%%%%%%%%%%%%%%%%%%%%%%%%%%%%%%%%%%%%%%%%%%%%%%%%%%%

\item {\bf [$\NS$-$\R$ ($\R$-$\NS$) sector ] }

Again, it is obvious that the right mover should correspond to a Ramond vacuum 
in the $SL(2,\br)/U(1)$-sector, and the lightest state has the conformal weights 
$h=\tilde{h} = \frac{1}{2} + \frac{1}{4k} $ due to the level-matching condition. 
It is explicitly realized by setting\footnote
   {Recall that we assumed $K\in 2\bz_{\geq 0}$.
Thus, such $v$ is always an integer.} 
$v=\frac{K}{2}$ 
$a = - \frac{1}{2}$ and $\ta=0$, which satisfies 
\begin{equation}
%v+ K(\nu+\tnu) =0,
v+K(a+\ta)=0, \hspace{1cm} v=0,\ldots, N-1, ~~ a \in \frac{1}{2} + \bz_N, ~~ \ta \in \bz_N.
\label{cond NSR}
\end{equation}

\end{list}

%%%%%%%%%%%%%%%%%%%%%%%%%%%%%%%%%%%%%%%%%%%%%%%%%%%%%%%%%%%%%%%%%%%%%%%%%%%%%%%%%%%%%%%

~

%\vspace{5mm}

\item {\bf $Z^{(\con)}_{0, [\sigma_L,\sigma_R]} (\tau) $-part :}  ~~
%%%

We have to be a little careful in order to evaluate the contribution of this sector.
All the NS-NS states are again not necessarily GSO projected due to the factors 
$
\frac{1}{1+(-1)^{t(\sigma_L)}q^{-\nu}}
$,
$
\left[ \frac{1}{1+(-1)^{t(\sigma_R)}q^{\tnu}} \right]^*
$
appearing in \eqn{Z con 0 general}. However,  these wrong GSO terms have the minimal conformal weight equal to 
$\frac{1}{2}+ \frac{1}{4k}$. In fact, if focusing on primary states, we find that the leading terms with the wrong GSO projection
always appear in the form of $(-1)^{t(\sigma_L)} q^{|\nu|}$ or $(-1)^{t(\sigma_R)} \left[ q^{|\tnu|}\right]^*$
and $|\nu|, |\tnu| \geq \frac{1}{2}$ holds.

Moreover, 
all of the correctly GSO projected NS-NS states as well as the R-R, NS-R (R-NS) states are  
massive since they obviously satisfy 
$$
h, \, \tilde{h} > \frac{1}{2} + \frac{1}{4k},
$$
as in $\hZ^{(\asp)}_{[\sigma_L,\sigma_R]}(\tau)$.

\end{itemize}

~

%%%%%%%%%%%%%%%%%%%%%%%%%%%%%%%%%%%%%%%%
%%%%%%%%%%%%%%%%%%%%%%%%%%%%%%%%%%%%%%%%

In conclusion, the no winding spectrum read off from 
$Z^{(\fin)}_{[\sigma_L,\sigma_R]}(\tau)$ includes a unique massless state 
and no tachyon in the NS-NS sector, 
whereas all the states with other spin structures are massive (lying above the mass gap). 
This result is quite expected: 
we have a unique NS-NS modulus with no winding that 
should be identified with the parameter 
of deviation from extremality $\mu$ \eqn{mu def}. 
The absence of corresponding R-R and NS-R moduli implies that 
it corresponds to the marginal deformation breaking the space-time SUSY. 
We also note that all the familiar moduli of relative distances among NS5-branes 
appearing in the BPS solution should be lifted up, and such deformations are absent in 
our near-extremal background \eqn{sol black NS5}. 
This fact is again consistent with our analysis of spectrum.

%\vspace{5mm} 

~

%\newpage
%%%%%%%%%%%%%%%%%%%%%%%%%%%%%%%%%%%%%%%%%%%%%%%%%%%%%%%%%%%%
%%%%%%%%%%%%%%%%%%%%%%%%%%%%%%%%%%%%%%%%%%%%%%%%%%%%%%%%%%%%

\noindent
{\bf (ii) thermal winding spectrum - `effective Hagedorn behavior'F}

We next discuss the thermal winding spectrum which captures the thermodynamical feature  
of the finite part $Z^{(\fin)}(\tau)$ according to the standard treatment of thermal string theory
\cite{thermal string,AW}. 
In other words, one will be aware of the UV behavior of the free energy of superstring gas by observing 
the spectrum with non-vanishing winding number $m_1\neq 0$ with the help of modular transformation, 
even though the winding states themselves are not regarded as physical ones.

For our purpose
it is enough to analyze the sector with winding number $m_1=1$, that is, 
with the constraint:
\begin{equation}
v+K(\nu+\tnu) = N, 
\label{cond thermal tachyon 0}
\end{equation}
since it obviously yields the leading contribution.

Similarly to the analysis of no winding spectrum, one can confirm that 
all the states satisfying \eqn{cond thermal tachyon 0} in 
$Z^{(\con)}_{0,[\sigma_L, \sigma_R]}$ have conformal weights greater than $\frac{k}{4}$,
although a part of terms are again wrong GSO-projected. 
It is quite anticipated and leads to the same thermal behavior 
as that of $\hZ^{(\asp)}_{[\sigma_L,\sigma_R]}(\tau)$ 
\eqn{hZ asp} consistent with the Hawking temperature \eqn{Hawking T}.

However, $Z^{(\dis)}_{0,[\sigma_L, \sigma_R]}$ yields more interesting aspects. 
We can again find a unique massless winding state in the NS-NS sector. 
In fact, it is sufficient to look for the lightest vacuum state lying in the spectral flow orbit 
%(in the discrete extended character $\chids{\sNS}$, in other words) 
satisfying  
\begin{equation}
v+K(a+\ta) =N, \hspace{1cm} v=0,\ldots, N-1,  ~~ a,\ta \in \frac{1}{2}+\bz_N,
\label{cond NSNS 2}
\end{equation}
in the NS-NS sector of \eqn{Z dis 0 general}.
The expected lightest state is obtained by setting $v=N-K$, $a=\ta = \frac{1}{2}$ (chiral primary).
This is wrong GSO projected due to the thermal phase factor $\bep(\sigma, *)$ \eqn{bep},  
and has the conformal weights
$h=\tilde{h} = \frac{1}{2}$ \footnote{
It would be worth pointing that
massless winding state presented here and 
the no winding one \eqn{no winding massless} 
are identified with the two Liouville potentials (screening charges) `$S^{\pm}$'
 in the dual picture
of $\cN=2$ Liouville theory \cite{FZZ2,LST,HK}.}. 
We also note that all the states appearing in the R-R, NS-R sectors with thermal winding 
have conformal weights greater than $\frac{k}{4}$, which again corresponds 
to the Hawking temperature.

To summarize, we have found a massless state with non-vanishing thermal winding  in the NS-NS sector. 
%which grows at most with a power law under the $\tau_2\, \rightarrow \, + \infty$ limit.  
This means that the physical excitations lying in $Z^{(\fin)}$ behave 
as if we were at the Hagedorn temperature, which is much higher than the Hawking temperature.

%%%%%%%%%%%%%%%%%%%%%%%%%%%%%%%%%%%%%%%%%%%%%%%%%%%%%%%%%%%%%%%%%%%%%%%%%%%%
%%%%%%%%%%%%%%%%%%%%%%%%%%%%%%%%%%%%%%%%%%%%%%%%%%%%%%%%%%%%%%%%%%%%%%%%%%%%
One could interpret this light excitation, from a geometrical viewpoint,   
as a contribution from the string wound around a small circle very close 
to the tip of cigar. 
However, it seems interesting to ask why we {\em exactly\/} observe the Hagedorn temperature.   
We would like to comment on a similarity to the effective Hagedorn behavior observed in the spectrum of closed string emission \cite{NST}
from the `rolling D-branes'  \cite{rolling D-brane}. 
In cases of \cite{NST}, the effective high temperature behavior would be interpreted as those 
caused by a very short `thermal' open string attached to points close to the tip 
of the `hairpin shaped D-brane' \cite{FZZT}, which is Wick rotated to the rolling D-brane.  
However,  similarly to the present situation, it is not so obvious why the Hagedorn temperature emerges precisely.

%%%%%%%%%%%%%%%%%%%%%%%%%%%%%%%%%%%%%%%%%%%%%%%%%%%%%%%%%%%%%%%%%%%%%%%%%%%%%
%%%%%%%%%%%%%%%%%%%%%%%%%%%%%%%%%%%%%%%%%%%%%%%%%%%%%%%%%%%%%%%%%%%%%%%%%%%%%

We finally point out that the `modular completed' partition function $Z^{(\dis)}$ \cite{ES-NH}
given in \eqn{Z reg decomp 2}, \eqn{Z dis general} independently shows 
this effective Hagedorn behavior, 
since it is modular invariant and contains the winding massless state considered above. 
This is actually the `minimum part' that has this property. 
Namely, the remaining function $Z^{(\reg)} - Z^{(\dis)}$ normally behaves in the UV-region  
consistent with the Hawking temperature \eqn{Hawking T}.

~

%%%%%%%%%%%%%%%%%%%%%%%%%%%%%%%%%%%%%%%%%%%%%%%%%%%%%%%%%%%%%%%%%%%%%%%%%%%%%%%%%%%%%%%%%%
%%%%%%%%%%%%%%%%%%%%%%%%%%%%%%%%%%%%%%%%%%%%%%%%%%%%%%%%%%%%%%%%%%%%%%%%%%%%%%%%%%%%%%%%%%
%%%%%%%%%%%%%%%%%%%%%%%%%%%%%%%%%%%%%%%%%%%%%%%%%%%%%%%%%%%%%%%%%%%%%%%%%%%%%%%%%%%%%%%%%%

\subsection{Spectrum in the $\bz_M$-orbifold}

Because of the translational invariance along 
the Euclidean time direction $t_E$, one may consider 
the $\bz_M$-orbifold of $SL(2;\br)/U(1)$-sector. 
By making this orbifolding, one will gain a deficit angle at the tip and 
the geometry gets singular, which would affect the analysis given above.
%As an application of analysis given above, 
Motivated by this expectation, let us explore the spectrum of light excitations in that background.  
We assume $N=ML$, $M,L\in \bz_{>0}$, $M\geq 2$ \footnote
  {It is always possible for an arbitrary positive integer $M$, since we do not assume $N$ and $K$ are coprime.},
and 
the regularized  partition function \eqn{thermal part fn 0} is replaced with 
%%%%%
\begin{eqnarray}
\hspace{-1cm}
Z^{ \msc{orb}, \, (\reg) }_{[\sigma_L,\sigma_R]}(\tau; \vep) & = &\frac{k}{M}  \sum_{m_i\in\bz}\,
\int_{\vep}^{1-\vep} ds_1 \int_0^1 ds_2 \,  \ep(\sigma_L; m_1,m_2) \ep(\sigma_R; m_1,m_2)\, 
\nn
&& ~~~ \times  
f_{[\sigma_L]}(\tau,s_1\tau+s_2)
\left[f_{[\sigma_R]}(\tau,s_1\tau+s_2)\right]^* \,
%\nn
%&& \hspace{1cm} \times
e^{-\frac{\pi k}{\tau_2}\left|\left(s_1+\frac{m_1}{M} \right)\tau+ \left(s_2+\frac{m_2}{M} \right) \right|^2} .
\label{orb thermal part fn 0}
\end{eqnarray}
We can likewise analyze this partition function and decompose it in the similar manner to 
\eqn{Z reg decomp}, \eqn{Z reg decomp 2} and \eqn{decomp asp fin}.
As is expected, it is found that the asymptotic part is precisely interpreted 
as the free superstring gas on the background
$$
S^1_{\beta_{\msc{Hw}}/M} \times \br_{\phi} \times SU(2)_k \times T^5,
$$
with the temperature $M T_{\msc{Hw}}$.

Let us turn our focus to the discrete part. 
By extending the analysis presented in \cite{ncpart-orb} so that the spin structures are included,
one can reach the following (modular completed) discrete partition function;
\begin{eqnarray}
Z^{\msc{orb},\, (\dis)}_{[\sigma_L,\sigma_R]}(\tau) &=& \sum_{v=0}^{N-1} \,
\sum_{\stackrel{(a,\ta) \in \cR(\sigma_L,\sigma_R,v;M) }{a \in \bz_N+ \frac{s(\sigma_L)-1}{2}, \, \ta \in \bz_N+ \frac{s(\sigma_R)-1}{2} }} \,
\bep(\sigma_L; v,a,\ta) \bep(\sigma_R; v,a,\ta) 
\nn
&&
\hspace{5mm}
\times
 \hchids{\sigma_L}(v, a;\tau) \, \overline{\hchids{\sigma_R}(v, \ta;\tau)} 
 \, \left(\frac{\th_{[\sigma_L]}}{\eta}\right)^3
\overline{ \left(\frac{\th_{[\sigma_R]}}{\eta}\right)^3}
\nn
%%%%%%%%%%%%
&=& 
\sum_{v=0}^{N-1} \,
\sum_{(\nu,\tnu) \in \cR(\sigma_L,\sigma_R,v;M)} \,
\bep(\sigma_L; v,a,\ta) \bep(\sigma_R; v,a,\ta) 
\nn
&&
\hspace{5mm}
\times
 \hchds{\sigma_L}\left(\frac{v}{K}, \nu ;\tau\right) \, \overline{\hchds{\sigma_R}\left(\frac{v}{K}, \tnu ;\tau\right)} 
 \, \left(\frac{\th_{[\sigma_L]}}{\eta}\right)^3
\overline{ \left(\frac{\th_{[\sigma_R]}}{\eta}\right)^3}, 
\label{orb Z dis}
\end{eqnarray}  
where the range of summation $\cR(\sigma_L,\sigma_R,v;M)$ has been defined as
\begin{eqnarray}
\hspace{-1.5cm}
&& \cR(\sigma_L,\sigma_R, v; M) := \left\{ (\nu ,\tnu )~| ~ \nu \in \bz + \frac{s(\sigma_L)-1}{2}, 
%\right.
%\nn
%&& 
%\hspace{0.5cm}
%\left. 
~ \tnu  \in \bz  + \frac{s(\sigma_R)-1}{2},
\right. 
\nn
&& \hspace{3cm}
\left.
~ v+K(\nu +\tnu) \in L\bz, ~ \nu -\tnu \in M \left(\bz+ \frac{s(\sigma_L)}{2} + \frac{s(\sigma_R)}{2} \right) \right\}.
\nn
&& 
\label{cR M}
\end{eqnarray}

Now, what spectrum will \eqn{orb Z dis} yield? The analysis becomes slightly complicated, and one has to be again careful 
about whether the GSO projection act correctly in the NS-sector.  
%%%%%%%%%%%%%%%%%%%%%%%%%%%%%%%%%%%%%%%%%%%%%%%%%%%%%%%%%%%%%%%%%%%%%%%%%%%%%%%%%%%%%%%%%%%%%%%%%%%%%
The results are summarized as follows;

\begin{description}
\item[(i) no winding spectrum:]

~

As in the previous analysis, 
we have a unique massless state and no tachyons in the NS-NS sector, 
and all the states with other spin structures are found to be massive. 
This is just given  by \eqn{no winding massless}, which clearly belongs to the range \eqn{cR M} for an arbitrary value of $M$. 
%\begin{equation}
%v=K, ~~~ a=\ta \equiv -\frac{1}{2} ~ (\mod\, N) ,
%\label{no winding massless 2}
%\end{equation} 
Consequently, we have a universal massless excitation irrespective of the orbifolding.
It should be identified with the modulus $\mu$ \eqn{mu def}.

%%%%%%%%%%%%%%%%%%%%%%%%%%%%%%%%%%%%%%%%

~

\item[(ii) thermal winding spectrum:]

~

We note that the thermal winding number $w \in \bz$ is now identified as 
\begin{equation}
v+ K (\nu +\tnu) = L w \left(\equiv \frac{N}{M} w \right).
\label{orb thermal winding} 
\end{equation}
Taking account of how the correct/wrong GSO projections act, 
we obtain the unique lightest excitation 
\begin{equation}
v= L-K, \hspace{1cm}  \nu = \tnu = -\frac{1}{2},
\label{orb thermal tachyon}
\end{equation}
which belongs to the NS-NS sector with the winding  $w=1$.
This state is wrong GSO projected and possesses the conformal weights\footnote
   {It might be worthwhile to comment on the following fact: contrary to the 
unorbifolded case, the R-R vacua  exist also 
in the winding sectors with $w=1,\ldots, M-1$.
They have the equal conformal weight $h=\tilde{h} = \frac{1}{2}+ \frac{1}{4k}$, which are massive but smaller than the weight for the thermal tachyon 
in the asymptotic region $\frac{1}{M^2} \frac{k}{4}$ as long as $M$ 
is sufficiently small. 
On the other hand, 
all the R-R winding states in the unorbifolded background have the weight greater 
than $\frac{k}{4}$. }
\begin{equation}
h = \tilde{h} = \frac{L}{2N} = \frac{1}{2M} < \frac{1}{2}. 
\label{orb thermal tachyon h} 
\end{equation}
Thus, it is  tachyonic. 
It is also not difficult to show that all the other states that belong to the partition function \eqn{orb Z dis} are massive.

In this way, we have now achieved the Hagedorn-like behavior again, but {\em above\/} the Hagedorn temperature.
Namely, a tachyonic instability is caused by the thermal winding states.  
Note that $L \geq K$ has to be satisfied for the thermal tachyon \eqn{orb thermal tachyon} to exist.
This constraint is equivalent with $M \leq k$, which is generic enough since we assumed a large $k$.

Emergence of such a tachyonic excitation would reflects  
the fact that the background contains a singularity.
%%%%%%%%%%%%%%%%%
It is also  interesting that the conformal weight \eqn{orb thermal tachyon h} 
is proportional to $M^{-1}$, rather than $M^{-2}$, in contrast with 
the thermal tachyon in the asymptotic sector, which  has the conformal weight 
$$
h = \tilde{h} = \frac{1}{M^2} \cdot \frac{k}{4} \equiv \left(\frac{1}{M} \frac{\beta_{\msc{Hw}}}{2\pi} \right)^2.
$$

\end{description}

~

We finally comment on the remaining sector $Z^{\msc{orb}, \, (\fin)} - Z^{\msc{orb}, \, (\dis)}$. 
Again this sector does not alter the relevant behavior of partition function, although the GSO-projection acts intricately:
states with no winding in this sector have conformal weights greater than  $\frac{1}{2} + \frac{1}{4k}$, 
while the conformal weights of winding states always satisfy the inequality
$h \geq \frac{1}{M^2} \frac{k}{4}$.

~

%%%%%%%%%%%%%%%%%%%%%%%%%%%%%%%%%%%%%%%%%%%%%%%%%%%%%%%%%%%%%%%%%%%%%%%%%%%%%%%%%%%%%%%%%%%%%%%%%%%%%%%
%%%%%%%%%%%%%%%%%%%%%%%%%%%%%%%%%%%%%%%%%%%%%%%%%%%%%%%%%%%%%%%%%%%%%%%%%%%%%%%%%%%%%%%%%%%%%%%%%%%%%%%
%%%%%%%%%%%%%%%%%%%%%%%%%%%%%%%%%%%%%%%%%%%%%%%%%%%%%%%%%%%%%%%%%%%%%%%%%%%%%%%%%%%%%%%%%%%%%%%%%%%%%%%

\section{Summary and Comments}

In this paper we have  studied  the thermal torus partition function of superstring propagating in the near-horizon region of
the near-extremal black NS5-brane background,  which is described by the superconformal system \eqn{SCFT NH black NS5}.

Main results are summarized as follows:
%%%%
\begin{itemize}
\item
The thermal partition function has been decomposed 
into two parts;
$$
Z^{(\reg)}(\tau) = Z^{(\asp)}(\tau) + Z^{(\fin)} (\tau),
$$
which we called, the `asymptotic part' and the `finite part'.

%%%%%%%%%%

\item 
The asymptotic part $Z^{(\asp)}(\tau)$ is contributed from strings freely propagating in the region 
far from the NS5-branes. 
%This sector includes a logarithmically divergent volume factor 
This part is written in  the same form as  the free superstring gas 
at the Hawking temperature in the linear-dilaton background, as already given in 
\cite{KutSah}. 
It includes contributions from free fermions with the correct thermal boundary conditions consistent 
with the GSO projection, when making the Wick rotation as given in \cite{AW}.

%%%%%%%%%%%

\item 
The finite part $Z^{(\fin)}(\tau)$ is a novel result of this paper. 
%which has been not given in 
%closely related studies, say, \cite{Mal-NENS5,MS-NENS5,BR,HO,KutSah}. 
It captures the contribution of strings localized around the `tip of cigar', 
which includes states not necessarily  GSO projected in the usual sense, 
and thus break the space-time SUSY even in the Lorentzian background. 
%characterizes the non-extremality.
This part only includes two massless states as the lightest state
in the NS-NS sector, and 
all the excitations with other spin structures are found to be massive.   
One of the massless excitations is identified 
as the energy excess above the extremality.
%(the value of dilaton at the tip of the cigar, in other words). 

Another massless state carries  a non-vanishing thermal winding. This fact implies that the Hagedorn-like behavior is effectively observed, 
although the Hawking temperature is much lower than the Hagedorn one. 
The `minimum part' that shows this effective Hagedorn behavior 
is identified  as  the discrete part $Z^{(\dis)} (\tau)$ \eqn{Z dis general}, 
written in terms of the modular completions \eqn{hchid}.

%This fact would be consistent with the thermal instability 
%in the Little String Theory (LST) \cite{LST} describing the large $N$ near-extremal NS5-branes
%discussed in \cite{BR,HO,KutSah}. 

The $\bz_M$-orbifold along the Euclidean time direction  has been 
similarly analyzed. The NS-NS modulus with no winding universally exists 
irrespective of the orbifolding. 
The effective Hagedorn behavior still happens. However, 
we this time observe a temperature higher than the Hagedorn one \eqn{Hagedorn T}:
the lightest thermal winding state becomes tachyonic, 
which would reflect the singularity of background.

\end{itemize}

~

%%%%%%%%%%%%%%%%%%%%%%%%%%%%%%%%%%%%%%%%%%%%%%%%%%%%%%%%%%%%%%%%%%%%%%%%%%%%%%%%%%%
%%%%%%%%%%%%%%%%%%%%%%%%%%%%%%%%%%%%%%%%%%%%%%%%%%%%%%%%%%%%%%%%%%%%%%%%%%%%%%%%%%%

We add several comments:

\begin{description}
\item[1.]
%\noindent
%{\bf 1.} 
Based on the standard argument of thermal string theory \cite{thermal string,AW}, 
the thermal partition function is reinterpreted as  the free energy of superstring gas
by setting $m_1(\equiv w) =0$ and replacing the fundamental region $\cF$ with the strip $\cS$ \eqn{strip} \cite{Polchinski}.
However, one should {\em not\/} suppose that the massless excitation with no winding we found belongs to the physical Hilbert space of single string,
after going back to the Lorentzian background. 
When translating the torus partition function into the free energy of multi-string systems, 
each sector with the temporal winding $m_2$ is identified as the contribution from the physical 
Hilbert space of $m_2$-string states, and recall that the temporal winding $m_2$ has been dualized into the KK momentum $n$ 
in our analysis. Therefore, the massless excitation we found would
rather correspond to a  {\em collective\/}  excitation of superstring gas.

%%%%%%%%%%%%%%%%%%%%%%%%%%%%%%%%%%%%%%%%%

\item[2.]
%\noindent
%{\bf 2.}  
It has been believed that the NS5-brane systems should be described by the Little String Theory (LST) \cite{LST} after taking  a suitable 
decoupling limit, which is assumed to be a 6-dimensional non-perturbative and non-gravitational string theory.
The fundamental superstring on the background \eqn{Euclidean NH black NS5} 
%we concentrate on in this paper 
is interpreted as the holographic dual of the thermal LST.
The Hagedorn temperature of LST  has been claimed to be equal the Hawking temperature $T_{\msc{Hw}}$ \eqn{Hawking T}
\cite{Mal-NENS5} based on the argument of microscopic origin of 
near-extremal black-hole entropy. 
Along this line, the Hagedorn behavior of LST has been investigated in \cite{BR,HO,KutSah}.
%\begin{equation}
%T_{\msc{Hg}}^{[\msc{LST}]}
%\end{equation}
%$$
%T_{\msc{Hg}}^{[\msc{LST}]} = T_{\msc{Hw}} \equiv 1/ \left(2\pi \sqrt{\al' k} \right).
%$$
%Although our main focus lies in the fundamental superstring, 
%we will later make a small comment on the relation of our analysis to 
%the Hagedorn behavior of LST.  
Especially, in the paper \cite{KutSah}, the perturbative region of 
fundamental superstring 
(that is, \eqn{cond weak coupling} is assumed)
%large $\mu$) 
has been claimed 
to be dual to the LST slightly above $T^{(\msc{LST})}_{\msc{Hg}} \left(\equiv T_{\msc{Hw}}\right)$,
and thus a thermal instability should emerge. 
The effective Hagedorn behavior observed in $Z^{(\fin)}(\tau)$ seems to be 
consistent with this claim.

%%%%%%%%%%%%%%%%%%%%%%%%%%%%%%%%%%%%%%%%%%%%%%%

\item[3.]
As we discussed in  section 4, 
the most relevant part for the effective Hagedorn behavior is 
the discrete partition function
$Z^{(\dis)}$ \eqn{Z dis general} that is modular invariant. 
%and the identity \eqn{id Z dis w} plays a crucial role to understand it. 
Recall that the modular completion $\hchids{\sigma}$ \eqn{hchid} roughly has the structure
\begin{equation}
\hchids{\sigma} = \chids{\sigma} + \sum \, \mbox{[non-holomorphic, massive]}.
\label{structure hchid}
\end{equation}
Because of the modular property of $\hchids{\sigma}$ \eqn{S hchid} (see also \eqn{S chid})
and the fact that the modular S-transformation exchanges the UV and IR regions in the torus moduli space, 
%the UV-region $\tau\, \sim \, +0$ with the IR-region $\tau \, \sim \, i \infty$, 
one can conclude that the  second term 
%(``shadow part'') 
in \eqn{structure hchid} should dominantly contribute in the UV region. 
While this term just provides a small correction to the discrete character $\chids{\sigma}$
in the IR region, 
{\em it dominates under the UV limit $\tau\, \rightarrow \, +0$,}  since 
it includes much more terms in its $q$-expansion than those of $\chids{\sigma}$. 
The accumulation of high energy excitations appearing there 
could affect the UV behavior of the partition function, 
leading to a thermal behavior typical at the Hagedorn temperature.

\end{description}

%%%%%%%%%%%%%%%%%%%%%%%%%%%%%%%%%%%%%%%%%%%%%%%%%%%%%%%%%%%%%%%%%%%%%%%%%%%%%%
%%%%%%%%%%%%%%%%%%%%%%%%%%%%%%%%%%%%%%%%%%%%%%%%%%%%%%%%%%%%%%%%%%%%%%%%%%%%%%
%%%%%%%%%%%%%%%%%%%%%%%%%%%%%%%%%%%%%%%%%%%%%%%%%%%%%%%%%%%%%%%%%%%%%%%%%%%%%%

~

%%%%%%%%%%%%%%%%%%%%%%%%%%%%%%%%%%%%%%%%%%%%%%%%%%%%%%%%%%%%%%%%%%%%
%%%%%%%%%%%%%%%%%%%%%%%%%%%%%%%%%%%%%%%%%%%%%%%%%%%%%%%%%%%%%%%%%%%%

\section*{Acknowledgments}

This work was supported by JSPS KAKENHI Grant Number 23540322 
from Japan Society for the Promotion of Science (JSPS).

%%%%%%%%%%%%%%%%%%%%%%%%%%%%%%%%%%%%%%%%%%%%%%%%%%%%%%%%%%%%%%%%%%%%

%~

%~

%%%%%%%%%%%%%%%%%%%%%%%%%%%%%%%%%%%%%%%%%%%%%%%%%%%%%%%%%%%%%%%%%%%%%%%%%%%%%%
%%%%%%%%%%%%%%%%%%%%%%%%%%%%%%%%%%%%%%%%%%%%%%%%%%%%%%%%%%%%%%%%%%%%%%%%%%%%%%
%%%%%%%%%%%%%%%%%%%%%%%%%%%%%%%%%%%%%%%%%%%%%%%%%%%%%%%%%%%%%%%%%%%%%%%%%%%%%%

\newpage

\section*{Appendix A: ~ Conventions for Theta Functions}

\setcounter{equation}{0}
\def\theequation{A.\arabic{equation}}

We assume $\tau\equiv \tau_1+i\tau_2$, $\tau_2>0$ and 
 set $q:= e^{2\pi i \tau}$, $y:=e^{2\pi i z}$;
 \begin{equation}
 \begin{array}{l}
 \dsp \th_1(\tau,z)=i\sum_{n=-\infty}^{\infty}(-1)^n q^{(n-1/2)^2/2} y^{n-1/2}
  \equiv 2 \sin(\pi z)q^{1/8}\prod_{m=1}^{\infty}
    (1-q^m)(1-yq^m)(1-y^{-1}q^m), \\
 \dsp \th_2(\tau,z)=\sum_{n=-\infty}^{\infty} q^{(n-1/2)^2/2} y^{n-1/2}
  \equiv 2 \cos(\pi z)q^{1/8}\prod_{m=1}^{\infty}
    (1-q^m)(1+yq^m)(1+y^{-1}q^m), \\
 \dsp \th_3(\tau,z)=\sum_{n=-\infty}^{\infty} q^{n^2/2} y^{n}
  \equiv \prod_{m=1}^{\infty}
    (1-q^m)(1+yq^{m-1/2})(1+y^{-1}q^{m-1/2}), \\
%\hspace{8cm} \mbox{(Jacobi's triple product identity)} \\
 \dsp \th_4(\tau,z)=\sum_{n=-\infty}^{\infty}(-1)^n q^{n^2/2} y^{n}
  \equiv \prod_{m=1}^{\infty}
    (1-q^m)(1-yq^{m-1/2})(1-y^{-1}q^{m-1/2}) .
 \end{array}
\label{th}
 \end{equation}
 \begin{eqnarray}
 \Th{m}{k}(\tau,z)&=&\sum_{n=-\infty}^{\infty}
 q^{k(n+\frac{m}{2k})^2}y^{k(n+\frac{m}{2k})} .
%\\
% \tTh{m}{k}(\tau,z)&=&\sum_{n=-\infty}^{\infty} (-1)^n
% q^{k(n+\frac{m}{2k})^2}y^{k(n+\frac{m}{2k})}.
 \end{eqnarray}
% We use abbreviations; $\th_i (\tau) \equiv \th_i(\tau, 0)$
% ($\th_1(\tau)\equiv 0$), 
%$\Th{m}{k}(\tau) \equiv \Th{m}{k}(\tau,0)$.
% $\tTh{m}{k}(\tau) \equiv \tTh{m}{k}(\tau,0)$.
 We also set
 \begin{equation}
 \eta(\tau)=q^{1/24}\prod_{n=1}^{\infty}(1-q^n).
 \end{equation}
%
%The anti-symmetrized theta functions 
%are defined as 
%\begin{eqnarray}
% && \Th{m}{k}^{(-)}(\tau,z) = \frac{1}{2}\left(
%\Th{m}{k}(\tau,z) - \Th{m}{k}(\tau,-z) 
%\right)\equiv \frac{1}{2}\left(
%\Th{m}{k}(\tau,z) - \Th{-m}{k}(\tau,z) 
%\right)~, \nn
% && \tTh{m}{k}^{(-)}(\tau,z) = \frac{1}{2}\left(
%\tTh{m}{k}(\tau,z) - \tTh{m}{k}(\tau,-z) 
%\right)\equiv \frac{1}{2}\left(
%\tTh{m}{k}(\tau,z) - \tTh{-m}{k}(\tau,z) 
%\right)~,
%\end{eqnarray}
% 
The spectral flow properties of theta functions are summarized 
as follows ($m,n, a \in \bz$, $k \in \bz_{>0}$);
\begin{eqnarray}
 && \th_1(\tau, z+m\tau+n) = (-1)^{m+n} 
q^{-\frac{m^2}{2}} y^{-m} \th_1(\tau,z) ~, \nn
&& \th_2(\tau, z+m\tau+n) = (-1)^{n} 
q^{-\frac{m^2}{2}} y^{-m} \th_2(\tau,z) ~, \nn
&& \th_3(\tau, z+m\tau+n) = 
q^{-\frac{m^2}{2}} y^{-m} \th_3(\tau,z) ~, \nn
&& \th_4(\tau, z+m\tau+n) = (-1)^{m} 
q^{-\frac{m^2}{2}} y^{-m} \th_4(\tau,z) ~, \nn
%%%
&& \Th{a}{k}(\tau, 2(z+m\tau+n)) = 
%e^{2\pi i n a} 
q^{-k m^2} y^{-2 k m} \Th{a+2km}{k}(\tau,2z)~.
%\nn
%&& \tTh{a}{k}(\tau,2(z+m\tau+n)) = (-1)^{m}e^{2\pi i n a} 
%q^{-k m^2 } y^{-2k m} \tTh{a+2km}{k}(\tau,2z)~,
\label{sflow theta}
\end{eqnarray}

%%%%%%%%%%%%%%%%%%%%%%%%%%%%%%%%%%%%%%%%%%%%%%%%%%%%%%%%%%%%
%%%%%%%%%%%%%%%%%%%%%%%%%%%%%%%%%%%%%%%%%%%%%%%%%%%%%%%%%%%%

We also use the following identities in the main text;
\begin{eqnarray}
\frac{\th_3(\tau,u)}{\th_1(\tau,u)} & =&
\frac{\th_3(\tau,0)}{i\eta(\tau)^3} \, \sum_{n\in\bz} \,
\frac{e^{2\pi i u \left(n+\frac{1}{2}\right)}}{1+q^{n+\frac{1}{2}}},
\nn
%%%
\frac{\th_4(\tau,u)}{\th_1(\tau,u)} &=&
\frac{\th_4(\tau,0)}{i\eta(\tau)^3}  \, \sum_{n\in\bz} \,
\frac{e^{2\pi i u \left(n+\frac{1}{2}\right)}}{1-q^{n+\frac{1}{2}}},
\nn
%%%
\frac{\th_2(\tau,u)}{\th_1(\tau,u)} &=&
\frac{\th_2(\tau,0)}{i\eta(\tau)^3} \, \sum_{n\in\bz} \,
\frac{e^{2\pi i u n}}{1+q^{n}}.
\label{theta id spin str}
\end{eqnarray}
They hold for $u\equiv s_1 \tau+ s_2$, $0< s_1 <1$,
which are proven by using the identity given {\em e.g.} in \cite{ES-NH}.

%%%%%%%%%%%%%%%%%%%%%%%%%%%%%%%%%%%%%%%%%%%%%%%%%%%%%%%%%%%%%%%%%%%%%%%%%%%%%
%%%%%%%%%%%%%%%%%%%%%%%%%%%%%%%%%%%%%%%%%%%%%%%%%%%%%%%%%%%%%%%%%%%%%%%%%%%%%

~

%%%%%%%%%%%%%%%%%%%%%%%%%%%%%%%%%%%%%%%%%%%%%%%%%%%%%%%%%%%%
%%%%%%%%%%%%%%%%%%%%%%%%%%%%%%%%%%%%%%%%%%%%%%%%%%%%%%%%%%%%

\section*{Appendix B:~ Irreducible and Extended Characters and their Modular Completions
with General Spin Structures}

\setcounter{equation}{0}
\def\theequation{C.\arabic{equation}}

%%%%%%%%%%%%%%%%%%%%%%%%%%%%%%%%%%%%%%%%%%%%%%%%%%%%%%%%%%%%%%%%%%%%%%%%%

In this appendix we summarize the definitions as well as 
useful formulas for the (extended) characters 
and their modular completions of the $\cN=2$ superconformal algebra with 
$\hc \left(\equiv \frac{c}{3} \right)= 1+ \frac{2}{k}$.
We shall include here the formulas with general spin structures extending those given in \cite{ES-NH,ncpart-orb}.
We assume $k= N/K$, $N,K\in \bz_{>0}$ (not assumed to be coprime), when considering the extended characters.

%We focus only on  the $\tR$-sector\footnote
%   {In this paper 
%we shall use the convention of $\tR$-characters 
%with the inverse sign compared to those of \cite{ES-NH,ES-BH,ES-C}, 
%so that the Witten indices appear with the positive sign. 
%(See \eqn{WI} below.)}, and when treating the extended characters, 
%we assume $k= N/K$, $(N,K \in \bz_{>0}) $ (but, not assume $N$ and $K$ are co-prime). 

~

%%%%%%%%%%%%%%%%%%%%%%%%%%%%%%%%%%%%%%%%%%%%%%%%%%%%%%%%%%%%%%%%%%%%%%%%%%%%%%%%%%%%%%%%%%
%%%%%%%%%%%%%%%%%%%%%%%%%%%%%%%%%%%%%%%%%%%%%%%%%%%%%%%%%%%%%%%%%%%%%%%%%%%%%%%%%%%%%%%%%%

To express spin structures concisely, 
we shall use the notation;
\begin{equation}
\th_{[\sigma]}(\tau,z) : = \th_3(\tau,z), ~ \th_4(\tau,z), ~ \th_2(\tau,z), ~ -i \th_1(\tau,z),
\label{thsigma}
\end{equation}
for $\sigma = \NS,~\tNS, ~ \R, ~\tR$ respectively.
We also set 
\begin{equation}
s(\sigma) := \left\{
\begin{array}{ll}
0 & ~~~ \sigma=\NS, ~ \tNS, \\
1 & ~~~ \sigma = \R, ~ \tR
\end{array}
\right.
%%%
\hspace{1cm}
t(\sigma) := \left\{
\begin{array}{ll}
0 & ~~~ \sigma=\NS, ~ \R, \\
1 & ~~~ \sigma = \tNS, ~ \tR
\end{array}
\right.
\nonumber
\end{equation}
and
\begin{equation}
\kappa (\sigma) :=  
\left\{
\begin{array}{ll}
1 & ~~~ \sigma=\NS, ~ \tNS, ~ \R \\
-i & ~~~ \sigma = \tR
\end{array}
\right.
\nonumber
\end{equation}
%%%%%%%%%%%%%%%%%%%%%%%%%%%%%%%%%%%%%%%%
It is also convenient to introduce the notations;
\begin{equation}
S\cdot \NS = \NS, ~~~ S \cdot \tNS = \R, ~~~ S \cdot \R = \tNS, ~~~ S\cdot \tR= \tR,
\end{equation}
%%%
\begin{equation}
T\cdot \NS = \tNS, ~~~ T \cdot \tNS = \NS, ~~~ T \cdot \R = \R, ~~~ T \cdot \tR= \tR,
\end{equation}
to write down the modular transformation formulas.

%%%%%%%%%%%%%%%%%%%%%%%%%%%%%%%%%%%%%%%%%%%%%%%%%%%%%%%%%%%%%%%%%%%%%%%%%%%%%%%%%%%%%%%%%%
%%%%%%%%%%%%%%%%%%%%%%%%%%%%%%%%%%%%%%%%%%%%%%%%%%%%%%%%%%%%%%%%%%%%%%%%%%%%%%%%%%%%%%%%%%

~

\noindent
{\bf \underline{Continuous (non-BPS) Characters}:}
%%%
\begin{equation}
\ch {(\sigma)}{} (P,\mu;\tau,z) := q^{\frac{P^2+ \mu^2}{4k}} 
y^{\frac{\mu}{k}}  \,
\frac{\th_{[\sigma]}(\tau,z)}{\eta(\tau)^3},
\label{ch c}
\end{equation}
which is associated to the irrep. with the following conformal weight $h$ 
and $U(1)$-charge $Q$; 
\begin{equation}
\begin{array}{lll}
\dsp h= \frac{P^2+ \mu^2}{4k}+\frac{1}{4k}, & \dsp ~ Q= \frac{\mu}{k} , &
~ (\mbox{for} ~ \sigma = \NS,~ \tNS) 
\\
\dsp h= \frac{P^2+ \mu^2}{4k}+\frac{\hc}{8}, & \dsp ~ Q= \frac{\mu}{k} \pm \frac{1}{2},
~(\mbox{doubly degenerated}) , &
~ (\mbox{for} ~ \sigma =  \R,~ \tR) 
\end{array}
\end{equation}

~

The modular transformation formulas 
and the spectral flow property 
are given by 
\begin{eqnarray}
&& 
\hspace{-1cm}
\ch{(\sigma)}{}\left(P,\mu ; - \frac{1}{\tau}, \frac{z}{\tau}\right)
= \kappa(\sigma)\, e^{i\pi \frac{\hc}{\tau}z^2}\, 
 \frac{1}{2k} \int_{-\infty}^{\infty} dP' \, 
\int_{-\infty}^{\infty} d\mu' \, 
e^{2\pi i \frac{P P' - \mu \mu'}{2k}}
\, \ch{(S\cdot\sigma)}{} (P',\mu';\tau,z) .
\label{S ch}
\\
&& \hspace{-1cm}
\ch{(\sigma)}{}\left(P,\mu ; \tau+1, z \right)
= e^{2\pi i \left(\frac{P^2+\mu^2}{4k} + \frac{s(\sigma)-1}{8}\right)}\,
\ch{(T\cdot \sigma)}{} \left(P, \mu ; \tau, z \right),
\label{T ch}
\\
%\end{eqnarray}
%%%%%%%%%%%%%%%%%%%%%%%%%%%%%%%%%%%%%%%%%%%%%%%%%%%%%%%%%%%%%
%\begin{equation}
&&
\hspace{-1cm}
\ch{(\sigma)}{} (P,\mu;\tau,z+n_1 \tau+n_2) = (-1)^{t(\sigma) n_1+s(\sigma) n_2} e^{2\pi i \frac{\mu}{k}n_2}
q^{-\frac{\hc}{2}n_1^2} y^{-\hc n_1}\, \ch{(\sigma)}{}(P,\mu+2n_1;\tau,z),
~~~ (\any n_i \in \bz). 
\nn
&&
\label{sflow ch}
\end{eqnarray}
%%%%%%%%%%%%%%%%%%%%%%%%%%%%%%%%%%%%%%%%%%%%%%%%%%%%%%%%%%%%%%%
%%%%%%%%%%%%%%%%%%%%%%%%%%%%%%%%%%%%%%%%%%%%%%%%%%%%%%%%%%%%%%%
%where we set 
%\begin{equation}
%S\cdot \NS = \NS, ~~~ S \cdot \tNS = \R, ~~~ S \cdot \R = \tNS, ~~~ S\cdot \tR= \tR,
%\end{equation}
%%%%
%\begin{equation}
%T\cdot \NS = \tNS, ~~~ T \cdot \tNS = \NS, ~~~ T \cdot \R = \R, ~~~ T \cdot \tR= \tR,
%\end{equation}
%and also 
%\begin{equation}
%\kappa (\sigma) :=  
%\left\{
%\begin{array}{ll}
%1 & ~~~ \sigma=\NS, ~ \tNS, ~ \R \\
%-i & ~~~ \sigma = \tR
%\end{array}
%\right.
%\end{equation}
%%%%%%%%%%%%%%%%%%%%%%%%%%%%%%%%%%%%%%%%%%%%%%%%%%%%%%%%%

~

\noindent
{\bf \underline{Discrete (BPS) Characters} \cite{BFK,Dobrev}:  }
%%%
\begin{eqnarray}
&& \chd^{(\sigma)} (\la,\nu;\tau,z) := \frac{(yq^{\nu})^{\frac{\la}{k}}}{1+ (-1)^{t(\sigma)}yq^{\nu}}\,
 y^{\frac{2\nu}{k}}  q^{\frac{\nu^2}{k}}\,
\frac{\th_{[\sigma]}(\tau,z)}{\eta(\tau)^3}, 
\nn
&& \hspace{6cm}
 \left( 0\leq \la \leq k, ~~ \nu \in  \bz+ \frac{s(\sigma)-1}{2} \right).
\label{ch d}
\end{eqnarray}
%%%%%%%%%%%%%%%%%%%
\begin{itemize}
%%%
\item 
For $\sigma = \NS, ~\tNS$ cases, this character 
is associated to the $\left(\nu-\frac{1}{2}\right)$-th spectral flow of 
discrete irrep. generated by the chiral primary with
\begin{equation}
h= \frac{Q}{2}= \frac{\la+1}{2k} , ~~~(0\leq \la \leq k). 
\end{equation}
%%%
\item
For $\sigma = \R, ~\tR$ cases, this character 
is associated to the $\nu$-th spectral flow of 
discrete irrep. generated by the Ramond vacua with 
%\footnote
%%%%%%%%%%%%%%%%%%%%%%%%%%%%%%%%
%%%%%%%%%%%%%%%%%%%%%%%%%%%%%%%%
%   {The unitarity requires $- \frac{\hc}{2} \leq Q \leq \frac{\hc}{2}$ for the Ramond 
%vacua, which is equivalent with the condition : $-1 \leq \la \leq k+1$ \cite{BFK}.
%The quantum number  $\la$ is identified with $2j-1$, where $j$ is the `isospin' 
%of $SL(2,\br)$ in the $SL(2,\br)/U(1)$-coset \cite{DPL}. Thus, the unitarity range  
%$-1 \leq \la \leq k+1$ corresponds to the `analogue of integrable representations'
%$
%0\leq j \leq \frac{k+2}{2}\equiv \frac{\kappa}{2},
%$
%where $\kappa$ denotes the level of bosonic $SL(2,\br)$-WZW. 
%The range $0\leq \la \leq k ~ (\Leftrightarrow ~ \frac{1}{2} \leq j \leq \frac{k+1}{2} )$ that we adopt here 
%is strictly narrower than this `unitarity range'.
%This restriction has a clear origin in the discrete spectrum of  the SUSY $SL(2,\br)/U(1)$-coset
%read off from the torus partition function. It is worth pointing out that this range 
%is invariant under modular transformations given below. We also note that the missing `edge' points
%$\la=-1,k+1$ correspond to the `graviton representation' and its spectral flows, which obey 
%different character formulas (see \cite{Dobrev}). 
%This type restriction of spectrum has been already discussed in 
%\cite{GK,MO,ES-BH}. };
%%%%%%%%%%%%%%%%%%%%%%%%%%%%%%%%%%
%%%%%%%%%%%%%%%%%%%%%%%%%%%%%%%%%%
\begin{equation}
h= \frac{\hc}{8}, ~~~ 
Q = \frac{\la}{k} - \frac{1}{2}, ~~~(0\leq \la \leq k) 
\end{equation}

\end{itemize}

~

%%%%%%%%%%%%%%%%%%%%%%%%%%%%%%%%%%%%%%%%%%%%%%%%%%%%%%%%%%%%%%
%%%%%%%%%%%%%%%%%%%%%%%%%%%%%%%%%%%%%%%%%%%%%%%%%%%%%%%%%%%%%%
The modular transformation formulas are given as \cite{ES-L} 
\begin{eqnarray}
&& 
\hspace{-1cm}
\chd^{(\sigma)} \left(\la,\nu ; - \frac{1}{\tau}, \frac{z}{\tau}\right)
= \kappa(\sigma) \, e^{i\pi \frac{\hc}{\tau}z^2}\,\left[
 \frac{i}{k} \int_0^k d\la'  \,\sum_{\nu' \in \bz+\frac{t(\sigma)-1}{2}}\,
e^{2\pi i \frac{\la \la' - (\la+2\nu)(\la'+2\nu')}{2k}}
\, \chd^{(S\cdot \sigma)} (\la',\nu';\tau,z) \right.
\nn
&& \hspace{5mm}
\left. + \frac{1}{2k} \, \int_{-\infty}^{\infty} d\mu'
\, e^{-2\pi i \frac{(\la+2\nu) \mu'}{2k}}\,\int_{\br(+i0)} dP'\,
 \frac{e^{-2\pi \frac{\la P'}{2k}}}{1+ (-1)^{s(\sigma)} e^{-\pi (P'+ i\mu')}}
\, \ch{(S\cdot \sigma)}{} (P',\mu';\tau,z) \right],
\label{S ch d}
\\
&& \hspace{-1cm}
\chd^{(\sigma)} \left(\la,\nu ; \tau+1, z \right)
= e^{2\pi i \left\{ \frac{\nu}{k} \left(\la+\nu \right)+ \frac{s(\sigma)-1}{8} \right\}}\,
\chd^{(T\cdot \sigma)} \left(\la,\nu ; \tau, z \right).
\label{T ch d}
\end{eqnarray}
%The spectral flow property is written as 
%\begin{equation}
%\chd (\la,n;\tau,z+r\tau+s) = (-1)^{r+s} e^{2\pi i \frac{\la+2n}{k}s}
%q^{-\frac{\hc}{2}r^2} y^{-\hc r}\, \chd(\la,n+r;\tau,z),
%~~~ (\any r,s \in \bz).
%\label{sflow chd}
%\end{equation}

~

%%%%%%%%%%%%%%%%%%%%%%%%%%%%%%%%%%%%%%%%%%%%%%%%%%%%%%%%%%%%%%%%%%%
%%%%%%%%%%%%%%%%%%%%%%%%%%%%%%%%%%%%%%%%%%%%%%%%%%%%%%%%%%%%%%%%%%%

\newpage

\noindent
{\bf \underline{Extended Continuous (non-BPS) Characters} \cite{ES-L,ES-BH}:}

Set $k= N/K$, $(N,K\in \bz_{>0})$.
%%%%%%%%%%%%%%%%%%%%%%%%%%%%%%%%%%%%%%%%%%%%%%%%%%%%%%%%%%%%%%%%%%%%%%%%%%%%%
\begin{eqnarray}
\chics{\sigma}(p,m;\tau,z) &:=  & \sum_{n\in N\bz}\, 
(-1)^{n t(\sigma)} q^{\frac{\hc}{2}n^2} y^{\hc n} \,
 \ch{(\sigma)}{}\left(\frac{p}{K}, \frac{m}{K}; \tau, z+n\tau\right)
\nn
&=& 
q^{\frac{p^2}{4NK}} \Th{m}{NK}\left(\tau,\frac{2z}{N}\right)\,
\frac{\th_{[\sigma]}(\tau,z)}{ \eta(\tau)^3}.
\label{chic}
\end{eqnarray}
%%%%%%%%%%%%%%%%%%%%%%%%%%%%%%%%%%%%%%%%%%%%%%%%%%%%%%%%%%%%%%%%%%%%%%%%%%%%%%
%This corresponds to the spectral flow sum of the non-degenerate representation with

$h= \frac{p^2+m^2+K^2}{4NK} + \frac{s(\sigma)}{8}$, 
$Q = \frac{m}{N}\pm \frac{s(\sigma)}{2}$
~($p\geq 0$, $m\in \bz_{2NK}$, doubly degenerated for $\sigma= \R,\,\tR$),

%whose flow momenta are taken to be $n\in N \bz$.
%%%%%%%%%%%%%%%%%%%%%%%%%%%%%%%%%%%%%%%%%%%%%%%%%%%%%%%%%%%%%%%%%%%%%%%%%%%%
The modular and spectral flow properties are simply written as 
\begin{eqnarray}
&& 
\hspace{-1cm}
\chics{\sigma}\left(p,m ; - \frac{1}{\tau}, \frac{z}{\tau}\right)
= \kappa(\sigma)\, e^{i\pi \frac{\hc}{\tau}z^2}\, 
 \frac{1}{2NK} \int_{-\infty}^{\infty} dp' \, 
\sum_{m'\in\bz_{2NK}}\, 
e^{2\pi i \frac{p p' - m m'}{2NK}}
\, \chics{S \cdot \sigma} (p',m';\tau,z).
\label{S chic}
\\
&& \hspace{-1cm}
\chics{\sigma} \left(p,m ; \tau+1, z \right)
= e^{2\pi i \frac{p^2+m^2}{4NK}+ \frac{s(\sigma)-1}{8}}\,
\chics{T\cdot \sigma} \left(p, m ; \tau, z \right),
\label{T chic}
\\
%\end{eqnarray}
%and the spectral flow property is given as 
%%%%%%%%%%%%%%%%%%%%%%%%%%%%%%%%%%%%%%%%%%%%%%%%%%%%%%%%%%%%%%
%\begin{equation}
&& \hspace{-1cm}
\chics{\sigma} (p,m;\tau,z+n_1\tau+n_2) = (-1)^{t(\sigma)n_1+s(\sigma)n_2} 
e^{2\pi i \frac{m}{N} n_2}
q^{-\frac{\hc}{2}{n_1}^2} y^{-\hc n_1}\, \chics{\sigma}(p,m+2Kn_1;\tau,z),
%~~~ (\any n_i \in \bz).
\nn
&&
\hspace{12cm} 
(\any n_i \in \bz).
\label{sflow chic}
\end{eqnarray}
%%%%%%%%%%%%%%%%%%%%%%%%%%%%%%%%%%%%%%%%%%%%%%%%%%%%%%%%%%%%%%%

~

%%%%%%%%%%%%%%%%%%%%%%%%%%%%%%%%%%%%%%%%%%%%%%%%%%%%%%%%%%%%
%%%%%%%%%%%%%%%%%%%%%%%%%%%%%%%%%%%%%%%%%%%%%%%%%%%%%%%%%%%%
%%%%%%%%%%%%%%%%%%%%%%%%%%%%%%%%%%%%%%%%%%%%%%%%%%%%%%%%%%%%

\noindent
{\bf \underline{Extended Discrete (BPS) Characters} \cite{ES-L,ES-BH,ES-C}:}
%%%%%%%%%%%%%%%%%%%%%%%%%%%%%%%%%%%%%%%%%%%%%%%%%%%%%%%%%%%%%%%%%%%
\begin{eqnarray}
\chids{\sigma} (v,a;\tau,z) &:= & 
 \sum_{n\in N\bz}\, 
(-1)^{nt(\sigma)} q^{\frac{\hc}{2}n^2} y^{\hc n} \, 
\chd^{(\sigma)}\left(\frac{v}{K}, a ; \tau, z+n\tau\right)
\nn
&=&  \sum_{n\in \bz}\,\chd^{(\sigma)}\left(\frac{v}{K}, a+N n ; \tau, z\right)
\nn
&=& \sum_{n\in\bz}\, 
\frac{(yq^{N n+ a})^{\frac{v}{N}}}
{1+(-1)^{t(\sigma)}yq^{Nn+a}} \, 
y^{2K\left(n+\frac{a}{N}\right)} q^{NK \left(n+\frac{a}{N}\right)^2}
\, \frac{\th_{[\sigma]}(\tau,z)}{\eta(\tau)^3}.
\label{chid}
\\
&& 
\hspace{2.5cm}
(v=0,1,\ldots, N-1, ~~ a \in \bz_N + \frac{s(\sigma)-1}{2}).
\nonumber
\end{eqnarray}
%%%%%%%%%%%%%%%%%%%%%%%%%%%%%%%%%%%%%%%%%%%%%%%%%%%%%%%%%%%%%%%%%%%%%%%
%%%%%%%%%%%%%%%%%%%%%%%%%%%%%%%%%%%%%%%%%%%%%%%%%%%%%%%%%%%%%%%%%%%%%%%

\begin{itemize}
\item $\sigma= \NS, \, \tNS$ :

This  corresponds to the  sum of the chiral primary representation 
with $h= \frac{1}{2}Q = \frac{v+K}{2N} $, 
~($v=0,1,\ldots , N-1$) over the spectral flows
 with flow momenta $m$  taken to be  
$m\in a-\frac{1}{2}+N\bz$, ($a\in \bz_N+\frac{1}{2}$).
%%%%%%%%%%%%%%%%%%%%%%%%%
Especially, in the case of $a= -\frac{1}{2} \left(\equiv N- \frac{1}{2}\right)$,
the corresponding spectral flow orbit is the one generated by 
the anti-chiral primary with 
$h= - \frac{1}{2}Q = \frac{N-v+K}{2N}$.

%%%%%%%%%%%%%%%%%%%%%%%

\item $\sigma = \R,\, \tR$ :

This  corresponds to the  sum of the 
Ramond vacuum representation with $h= \frac{\hc}{8}$, 
$Q= \frac{v}{N}-\frac{1}{2}$
~($v=0,1,\ldots , N-1$) over spectral flow
 with flow momentum $m$ 
taken to be  $m\in a +N\bz$.

\end{itemize}

%This again corresponds to the  sum of the 
%Ramond vacuum representation with $h= \frac{\hc}{8}$, 
%$Q= \frac{v}{N}-\frac{1}{2}$
%~($v=0,1,\ldots , N-1$) over spectral flow
% with flow momentum $m$  taken to be mod.$N$, as 
%$m= a +N\bz$ ~ ($a\in \frac{s(\sigma)-1}{2}+\bz_N$).
%%%%%%%%%%%%%%%%%%%%%%%%%%%%%%%%%%%%%%%%%%%%%%%%%%%%%%%%%%%%%%%%%%%%%%%
%%%%%%%%%%%%%%%%%%%%%%%%%%%%%%%%%%%%%%%%%%%%%%%%%%%%%%%%%%%%%%%%%%%%%%%
%If one introduces the notation of Appell function or Lerch sum
%\cite{Pol,STT,Zwegers}, 
%\begin{equation}
% \cK^{(2k)}(\tau,z) 
%:= \sum_{n\in \bsz} \frac{q^{kn^2} y^{2kn}}
%{1-yq^n}
%\label{Appell}
%\end{equation}
%one can write as
%\begin{eqnarray}
%&& \chid (v,a;\tau,z) = \frac{1}{N} \sum_{b\in\bz_N}\,
% e^{-2\pi i \frac{v b}{N}} q^{\frac{K}{N} a^2} y^{\frac{2K}{N} a}\,
% \cK^{(2NK)}\left(\tau, \frac{z+a\tau+b}{N}\right)\, 
%\frac{i\th_1(\tau,z)}{\eta(\tau)^3},
%\nn
%%%%
%&& q^{\frac{K}{N} a^2} y^{\frac{2K}{N} a}\,
% \cK^{(2NK)}\left(\tau, \frac{z+a\tau+b}{N}\right)\, 
%\frac{i\th_1(\tau,z)}{\eta(\tau)^3} 
%= \sum_{v=0}^{N-1} \,  e^{2\pi i \frac{v b}{N}} \, \chid (v,a;\tau,z).
%\label{rel chid cK} 
%\end{eqnarray}

The modular transformation formula 
% and $\cK^{(2k)}$ 
can be expressed as \cite{ES-L,ES-BH,ES-C};
%%%
\begin{eqnarray}
&& 
\hspace{-1cm}
\chids{\sigma} \left(v,a ; - \frac{1}{\tau}, \frac{z}{\tau}\right)
= \kappa(\sigma) e^{i\pi \frac{\hc}{\tau}z^2}\,\left[
 \sum_{v'=0}^{N-1} \,\sum_{a'\in \bz_N + \frac{t(\sigma)-1}{2}}\,
\frac{i}{N} \, e^{2\pi i \frac{vv' - (v+2Ka)(v'+2Ka')}{2NK}}
\, \chids{S\cdot\sigma}  (v',a';\tau,z) \right.
\nn
&& \hspace{0.5cm}
\left. +\frac{1}{2NK} \sum_{m' \in \bz_{2NK}} \, e^{-2\pi i \frac{(v+2Ka) m'}{2NK}}\,
\int_{\br+i0} dp'\, \frac{e^{-2\pi \frac{vp'}{2NK}}}
{1+(-1)^{s(\sigma)}e^{-\pi \frac{p'+im'}{K}}}
\, \chics{S\cdot \sigma} (p',m';\tau,z)
\right],
\label{S chid}
\\
%%%
&& 
\hspace{-1cm}
\chids{\sigma} \left(v,a ; \tau+1, z \right)
= e^{2\pi i \frac{a}{N} \left(v+ K a \right)}\,
\chids{T\cdot \sigma} \left(v,a ; \tau, z \right),
\label{T chid} 
\end{eqnarray}
%%%
%\begin{equation}
%\cK^{(2 k)}\left(-\frac{1}{\tau}, \frac{z}{\tau}\right)
%= \tau e^{i\pi  \frac{ 2k z^2}{\tau}}\,
%\left[ \cK^{(2k)}(\tau,z) - \frac{i}{\sqrt{2 k}}\, \sum_{m\in \bz_{2k}}\,
%\int_{\br+i0} dp' \, \frac{q^{\frac{1}{2}p^{'2}}}
%{1-e^{-2\pi \left(\frac{p'}{\sqrt{2k}}+i\frac{m}{2k}\right)}}\,
%\Th{m}{k}(\tau,2z)
%\right]
%\label{S cK}
%\end{equation}
%Integral over $p'$ in the above formulas  
%is called the Mordell's integral \cite{Mordell,Watson}.
%
%%%%%%%%%%%%%%%%%%%%%%%%%%%%%%%%%%%%%%%%%%%%%%%%%%%%%%%%%%%%%%%%%%%

The spectral flow property is also expressed as \cite{ES-C}
\begin{equation}
\chids{\sigma} (v,a;\tau,z+n_1\tau+n_2) = 
(-1)^{t(\sigma)n_1+s(\sigma)n_2} e^{2\pi i \frac{v+2Ka}{N}n_2}
q^{-\frac{\hc}{2}{n_1}^2} y^{-\hc n_1}\, \chids{\sigma}(v,a+n_1;\tau,z),
~~~ (\any n_i \in \bz),
\label{sflow chid}
\end{equation}

~

%\newpage

%%%%%%%%%%%%%%%%%%%%%%%%%%%%%%%%%%%%%%%%%%%%%%%%%%%%%%%%%%%%%%%%%%%%%%%%%%%%%%%
% Modular completions for general spin strucutures
%%%%%%%%%%%%%%%%%%%%%%%%%%%%%%%%%%%%%%%%%%%%%%%%%%%%%%%%%%%%%%%%%%%%%%%%%%%%%%%
\noindent
{\bf \underline{Modular Completion of the Irreducible Discrete Character} \cite{ncpart-orb}:}
%$(\nu \in \bz+\frac{s(\sigma)-1}{2})$ 
\begin{eqnarray}
&&
\hspace{-1cm}
\hchd^{(\sigma)}(\la,\nu;\tau,z) :=  
\chd^{(\sigma)}(\la,\nu;\tau,z) 
\nn
&& 
\hspace{1cm}
- \frac{1}{2} \sum_{r\in \bz}\, (-1)^{r\left(t(\sigma)-1\right)} 
\sgn(r +0)\, 
\erfc \left(\sqrt{\frac{\pi \tau_2}{k}} \left|\la+k r \right|\right)\, 
q^{\frac{\nu^2}{k}+ \frac{\nu}{k}(\la+kr)}\,
y^{\frac{1}{k} (\la+kr +2\nu)}\, \frac{\th_{[\sigma]}(\tau,z)}{\eta(\tau)^3} 
\nn
&& 
%\hspace{1cm}
=  \frac{\th_{[\sigma]}(\tau,z)}{\eta(\tau)^3} \, y^{\frac{2\nu}{k}}
q^{\frac{\nu^2}{k}}
\, \left\lb 
\frac{(yq^{\nu})^{\frac{\la}{k}}}{1+(-1)^{t(\sigma)}yq^\nu} 
\right.
\nn
&& \hspace{3cm} \left.
-\frac{1}{2} \sum_{r\in \bz}\, (-1)^{r\left(t(\sigma)-1\right)} 
\sgn(r +0)\, 
\erfc \left(\sqrt{\frac{\pi \tau_2}{k}} \left|\la+kr \right| \right)\, 
%q^{\frac{n \nu}{k}} \, y^{\frac{\nu}{k}}
(y q^\nu)^{\frac{\la+kr}{k}}
\right\rb
\nn
%%%
&&
%\hspace{1cm}
=   \frac{i \th_{[\sigma]}(\tau,z)}{ 2\pi \eta(\tau)^3}\,
\frac{y^{\frac{2\nu}{k}} q^{\frac{\nu^2}{k}}}{1+(-1)^{t(\sigma)}y q^{\nu}}
 \left\{ \int_{\br + i(k-0)} dp\, +(-1)^{t(\sigma)}
\int_{\br-i0} dp \, \left(y q^{\nu} \right) 
\right\}
\nn
&&
\hspace{3cm}
\times 
\sum_{r\in \bz}\, (-1)^{r\left(t(\sigma)-1\right)}
\frac{ e^{- \pi \tau_2 \frac{p^2+(\la+kr)^2}{k}} 
\left(y q^{\nu}\right)^{\frac{\la+kr}{k}}}{p-i(\la+kr)} ,
\nn
&& 
\hspace{8cm}
(0\leq \la \leq k, ~~\nu \in \bz+ \frac{s(\sigma)-1}{2}).
\label{hchd}
\end{eqnarray}
%$(0\leq \la \leq k, ~~n\in \bz)$,
In the non-holomorphic terms in \eqn{hchd}, 
 $\erfc(*)$ denotes the error-function defined by
\begin{equation}
\erfc(x):= \frac{2}{\sqrt{\pi}} \int_x^{\infty} e^{-t^2} \, dt
\left( \equiv 1- \erf(x) \right).
\label{erfc}
\end{equation}
The equality in the last line of \eqn{hchd} is derived from the integral formula;
\begin{equation}
\frac{1}{i\pi} \int_{\br\mp i0} dp\, \frac{e^{-\al (p^2+ \nu^2)}}{p-i\nu} = 
\sgn(\nu \pm 0) 
\erfc(\sqrt{\al}|\nu|), 
\hspace{1cm}(\nu \in \br, ~ \al >0),
\label{formula erfc}
\end{equation}
and by using a simple contour deformation technique.

%%%%%%%%%%%%%%%%%%%%%%%%
%Note that $\hchd(\la,n;\tau,z) $ is 
%non-holomorphic due to the explicit dependence on 
%$\tau_2 \equiv \Im \tau$.
%%%%%
%It is crucial that the modular completion $\hchd$ has nice modular properties. 
%Especially, one can prove that the S-transformation formula gets considerably simplified;
%%%%%%%%%%%%%%%%%%%%%%%

Now, the modular S-transformation formula is written as 
\begin{eqnarray}
&& 
\hspace{-1.5cm}
\hchd^{(\sigma)} \left(\la,\nu ; - \frac{1}{\tau}, \frac{z}{\tau}\right)
= \kappa(\sigma)
e^{i\pi \frac{\hc}{\tau}z^2}\,
 \frac{i}{k} \int_0^k d\la'  \,\sum_{\nu' \in \bz + \frac{t(\sigma)-1}{2}}\,
e^{2\pi i \frac{\la \la' - (\la+2\nu)(\la'+2\nu')}{2k}}
\, \hchd^{(S\cdot \sigma)} (\la',\nu';\tau,z).
\label{S hchd}
\end{eqnarray}
Namely, the continuous term appearing in the R.H.S of \eqn{S ch d} drops off by taking 
the modular completion, and the S-transformation formula gets closed within 
$\hchd^{(\sigma)}$.

On the other hand, 
the T-transformation and spectral flow property 
are preserved by taking the completion;
\begin{eqnarray}
&& 
\hchd^{(\sigma)} \left(\la,\nu ; \tau+1, z \right)
= e^{2\pi i \left\{ \frac{\nu}{k} \left(\la+ \nu \right)+ 
\frac{s(\sigma)-1}{8}\right\}}\,
\hchd^{(T\cdot \sigma)} \left(\la,n ; \tau, z \right),
\label{T hchd}
\\
%%%%%%%%%%%%%%%%%%%%%%%%%%%%%%%%%%%%%%%%%%%%%%%%%%%%%%%%%%%%%%%%%%%%%%%%
&& 
\hchd^{(\sigma)} (\la,\nu;\tau,z+n_1\tau+n_2) = (-1)^{t(\sigma)n_1+s(\sigma) n_2} 
e^{2\pi i \frac{\la+2\nu}{k}n_2}
q^{-\frac{\hc}{2}n_1^2} y^{-\hc n_1 }\, \hchd^{(\sigma)}(\la,\nu+n_1 ;\tau,z),
%~~~ (\any n_i \in \bz).
\nn
&& \hspace{12cm} (\any n_i \in \bz).
\label{sflow hchd}
\end{eqnarray}
%%%%%%%%%%%%%%%%%%%%%%%%%%%%%%%%%%%%%%%%%%%%%%%%%%%%%%%%%%%%%%%%%%%%%%%
%%%%%%%%%%%%%%%%%%%%%%%%%%%%%%%%%%%%%%%%%%%%%%%%%%%%%%%%%%%%%%%%%%%%%

%%%%%%%%%%%%%%%%%%%%%%%%%%%%%%%%%%%%%%%%%%%%%%%%%%%%%%%%%%
% charge conjugation relations
%%%%%%%%%%%%%%%%%%%%%%%%%%%%%%%%%%%%%%%%%%%%%%%%%%%%%%%%%%
%We note the following `charge conjugation relations';
%\begin{eqnarray}
%&& \hchds{\sigma}(\la, \nu ; \tau, -z) = \pm \hchds{\sigma} (k-\la, -\nu ; \tau, z), 
%\nn
%&& 
%\hspace{3cm}
%\left(+\,:\, \sigma = \NS, \, \R,\, \tR,  \hspace{1cm}  -\,:\, \sigma = \tNS\right),
%\label{CC hchd}
%\end{eqnarray}
%and also\footnote
%  {Note that the same type identity as \eqn{0 k id hchd} does not hold for the %irreducible characters $\chds{\sigma}$.  
%We instead obtain 
%$$
% \chds{\sigma}(k, \nu ; \tau, z) 
%= \pm \chds{\sigma} (0, \nu ; \tau, z) + \mbox{massive character}.
%$$
%}
%%%%%%%%%%%%%%%%%%%%%%%%%%%%%%%%%%%%%%%%%%%%%%%%%%%%%%%%%%%%%
%\begin{eqnarray}
%&& \hchds{\sigma}(k, \nu ; \tau, z) = \pm \hchds{\sigma} (0, \nu ; \tau, z), 
%\nn
%&& 
%\hspace{3cm}
%\left(+\,:\, \sigma = \tNS, \, \tR,  \hspace{1cm}  -\,:\, \sigma = \NS, \, \R \right).
%\label{0 k id hchd}
%\end{eqnarray}
%%%%%%%%%%%%%%%%%%%%%%%%%%%%%%%%%%%%%%%%%%%%%%%%%%%%%%%%%%%%%%%%%%%
%%%%%%%%%%%%%%%%%%%%%%%%%%%%%%%%%%%%%%%%%%%%%%%%%%%%%%%%%%%%%%%%%%%

~

%%%%%%%%%%%%%%%%%%%%%%%%%%%%%%%%%%%%%%%%%%%%%%%%%%%%%%%%%%%%%%%%%%
%%%%%%%%%%%%%%%%%%%%%%%%%%%%%%%%%%%%%%%%%%%%%%%%%%%%%%%%%%%%%%%%%%
%%%%%%%%%%%%%%%%%%%%%%%%%%%%%%%%%%%%%%%%%%%%%%%%%%%%%%%%%%%%%%%%%%

\noindent
{\bf \underline{Modular Completion of the Extended Discrete Characters} \cite{ES-NH}:}

The modular completion of the discrete character $\chids{\sigma}$ is defined as 
the spectral flow sum of $\hchds{\sigma}$ \eqn{hchd} in the similar manner to \eqn{chid};
\begin{eqnarray}
\hspace{-1cm}
\hchids{\sigma} (v,a;\tau,z) 
&:= & \sum_{n\in N\bz} \, (-1)^{nt(\sigma)} q^{\frac{\hc}{2}n^2} y^{\hc n}\,
\hchd^{(\sigma)}\left(\frac{v}{K}, a;\tau,z+n\tau\right)
\nn
&=&
 \sum_{m\in\bz}\, \hchd^{(\sigma)}\left(\frac{v}{K}, a + Nm ;\tau,z\right)
\nn
&=  &  \chids{\sigma} (v,a;\tau,z) - \frac{1}{2} \sum_{j\in \bz_{2K}}\,
(-1)^{j\left(t(\sigma)-1\right)}
R_{v+Nj, NK}(\tau) \Th{v+Nj+2Ka}{NK}\left(\tau, \frac{2z}{N}\right)\,
\frac{\th_{[\sigma]}(\tau,z)}{ \eta(\tau)^3},
\nn
%&\equiv & \frac{1}{N} \sum_{b\in\bz_N}\,
% e^{-2\pi i \frac{v b}{N}} q^{\frac{K}{N} a^2} y^{\frac{2K}{N} a}\,
% \hcK^{(2NK)}\left(\tau, \frac{z+a\tau+b}{N}\right)\, 
%\frac{i\th_1(\tau,z)}{\eta(\tau)^3}.
%%%%%%%%%%%%%%%%%%%%%%%%%%%%%%%%%%%%%%%%%%%
&=& 
\frac{i \th_{[\sigma]}(\tau,z)}{ 2\pi \eta(\tau)^3}\,
%\sum_{n\in a + N \bz}
\sum_{n,r\in\bz}\, \left[ (-1)^{r\left(t(\sigma)-1 \right)}
\frac{\left(y q^{Nn+a}\right)^{\frac{v+Nr}{N}}}
{1+(-1)^{t(\sigma)}y q^{Nn+a}} y^{2K \left(n+\frac{a}{N}\right)} q^{NK\left(n+\frac{a}{N}\right)^2}
\right.
\nn
&& \hspace{1.5cm}
\left.
\times 
\left\{ \int_{\br + i(N-0)} dp\, + (-1)^{t(\sigma)}
\int_{\br-i0} dp \, \left(y q^{Nn+a} \right) 
\right\}
\,
\frac{ e^{- \pi \tau_2 \frac{p^2+(Nn+a)^2}{NK}} }{p-i(v+Nr)} \right],
\nn
&&
\hspace{5cm} 
(v=0,1,\ldots, N, ~~~ a \in \bz_N+\frac{s(\sigma)-1}{2})
\label{hchid}
\end{eqnarray}
where we set 
\begin{eqnarray}
R_{m,k}(\tau) &:=& \sum_{\nu \in m+2k\bz}\, \sgn(\nu + 0) 
\erfc\left(\sqrt{\frac{\pi \tau_2}{k}} \left|\nu\right|\right)\, q^{- \frac{\nu^2}{4k}}
\nn
&=& \frac{1}{i\pi}\, \sum_{\nu \in m+2k\bz}\,
\int_{\br- i0} dp \, \frac{e^{-\pi \tau_2 \frac{p^2+\nu^2}{k}} }{p-i\nu}\,
q^{- \frac{\nu^2}{4k}}.
\label{Rmk}
\end{eqnarray}
%
% In (\ref{Rmk}) 
%it supplies a strong enough damping factor to make the power series convergent.
%Note that $\hcK^{(k)}(\tau,z)$ is holomorphic with respect to $z$, but {\em  not}
%with respect to $\tau$, since $R^{(\pm)}_{m,k}$ depends on $\tau_2$.  
%%%

Conversely the irreducible modular completion $\hchds{\sigma}$ \eqn{hchd}
is reconstructed from the extended one $\hchids{\sigma}$
\eqn{hchid} by taking the `continuum limit' \cite{ncpart-orb};
\begin{equation}
\lim_{\stackrel{N\,\rightarrow\, \infty}{k\equiv N/K }\, \msc{fixed}}\, 
\hchids{\sigma}\left(v,a;\tau,z\right) = 
\hchd^{(\sigma)} \left(\la \equiv \frac{v}{K}, a ;\tau,z\right).
\label{rel hchid hchd} 
\end{equation}

%%%%%%%%%%%%%%%%%%%%%%%%%%%%%%%%%%%%%%
%%%%%%%%%%%%%%%%%%%%%%%%%%%%%%%%%%%%%%
%%%%%%%%%%%%%%%%%%%%%%%%%%%%%%%%%%%%%%
%In the main text of this paper we also use an alternative notation
%\begin{eqnarray}
%&& 
%\bhchi(v,m;\tau,z) \equiv \hchid(v,a;\tau,z),  ~ ~  \mbox{with} ~ 
%m\equiv  v+2Ka \in \bz_{2NK}, ~~~ v=0, 1 \ldots, N-1,
%\nn
%%%%
%&& \bhchi(v,m;\tau,z) \equiv 0, ~~ \mbox{if} ~ m-v \not\in 2K \bz.
%\label{bhchi def}
%\end{eqnarray}
%%%%%%%%%%%%%%%%%%%%%%%%%%%%%%%%%%%%%%%
%%%%%%%%%%%%%%%%%%%%%%%%%%%%%%%%%%%%%%%
%%%%%%%%%%%%%%%%%%%%%%%%%%%%%%%%%%%%%%%

%The modular completion 
%$\hchid$ 
%%\eqn{hchid} 
%is non-holomorphic with respect to $\tau$, however, has good modular properties %\cite{ES-NH};
The modular transformation formulas for $\hchids{\sigma}$
\eqn{hchid} are written as 
\begin{eqnarray}
&& 
%\hspace{-1.5cm}
\hchids{\sigma} \left(v,a ; - \frac{1}{\tau}, \frac{z}{\tau}\right)
= \kappa(\sigma) e^{i\pi \frac{\hc}{\tau}z^2}\, 
\sum_{v'=0}^{N-1} \,\sum_{a'\in \bz_N+ \frac{t(\sigma)-1}{2}}\,
\frac{i}{N} \, e^{2\pi i \frac{vv' - (v+2Ka)(v'+2Ka')}{2NK}}
\, \hchids{S\cdot \sigma} (v',a';\tau,z),
\nn
&&
\label{S hchid}
\\
%%%
&& 
% \hspace{-1.5cm}
\hchids{\sigma} \left(v,a ; \tau+1, z \right)
= e^{2\pi i \left\{ \frac{a}{N} \left(v+ K a \right)+ \frac{s(\sigma)-1}{8}\right\}}\,
\hchids{T\cdot \sigma} \left(v,a ; \tau, z \right).
\label{T hchid}
\end{eqnarray}
Also the spectral flow property is preserved by taking the completion;
\begin{eqnarray}
&& \hspace{-5mm}
\hchids{\sigma} (v,a;\tau,z+n_1\tau+n_2) = (-1)^{t(\sigma)n_1+s(\sigma)n_2} 
e^{2\pi i \frac{v+2Ka}{N}n_2}
q^{-\frac{\hc}{2}n_1^2} y^{-\hc n_1}\, \hchids{\sigma}(v,a+n_1;\tau,z),
\nn
&& 
\hspace{12cm}
 (\any n_i \in \bz).
\label{sflow hchid}
\end{eqnarray}

\end{document}